\documentclass[amsmath,amssymb,prx,superscriptaddress,reprint,showpacs,longbibliography]{revtex4-1}
\usepackage{graphicx}
\usepackage{dcolumn}
\usepackage[colorlinks=true,linkcolor=blue,citecolor=blue,urlcolor=blue]{hyperref}
\usepackage[usenames,dvipsnames]{xcolor}
\usepackage{soul}
\usepackage{braket}
\usepackage{bm}
\usepackage{upgreek}
\usepackage{mathtools}

\begin{document}

\title{Vortex Lattice Structure and Topological Superconductivity \\ in the Quantum Hall Regime}

\author{Gaurav Chaudhary}
\email{gaurav-ph@utexas.edu}
\affiliation{Department of Physics, The University of Texas at Austin, Austin, Texas 78712, USA}
\author{Allan H. MacDonald}
\affiliation{Department of Physics, The University of Texas at Austin, Austin, Texas 78712, USA}

\date{\today}



\begin{abstract}
Chiral topological superconductors are expected to appear as intermediate states 
when a quantum anomalous Hall system is proximity coupled 
to an $s$-wave superconductor and the magnetization direction is reversed.  
In this paper we address the edge state properties of ordinary quantum Hall systems 
proximity coupled to $s$-wave superconductors, accounting explicitly for Landau quantization.
We find that the appearance of topological superconducting phases with an odd number of Majorana edge modes 
is dependent on the structure of the system's vortex lattice. 
More precisely, vortex lattices containing odd number of superconducting flux quanta per unit cell, 
always support an even number of chiral edge channels 
and are therefore adiabatically connected to normal quantum Hall insulators. 
We discuss strategies to engineer chiral topological superconductivity in proximity-coupled 
quantum Hall systems by manipulating vortex lattice structure.
\end{abstract}

\maketitle

\section{Introduction}\label{Sec:intro}

The quest for Majorana modes in two-dimensional electron systems 
has been one of the most active areas of condensed matter research over the past decade. Majorana modes were initially proposed by Ettore Majorana~\cite{Majorana1937,Majoranaa} as an interpretation of real solutions of the free particle Dirac equation in which particles are their own antiparticles.
The topic of Majorana particles and modes then 
received an interesting boost from condensed matter physics 
after the theoretical prediction of their presence as
quasiparticles of the $\nu = 5/2$ fractional quantum Hall state, and as 
zero energy Bogoliubov-de-Gennes (BdG) quasiparticles in superconductors with 
$p\pm ip$ pairing symmetry~\cite{Read2000,Ivanov2001}.

Two-dimensional (2D) superconductors with $p\pm ip$ pairing break time reversal ($\mathcal{T}$)-symmetry
and have low energy propagating chiral edge modes that 
satisfy the Majorana property $\gamma^{\dagger}_{\bm{k}} = \gamma_{-\bm{k}}$.
The appearance of edge localized quasiparticles in these superconductors is related to the topological classification of the 
Bogoliubov-de-Gennes (BdG) mean-field Hamiltonians that appear in the theory of superconductivity.
An important distinction between the topological classification of superconductors 
and that of ordinary non-interacting band electrons
is that the BdG Hamiltonian acts in a doubled Hilbert space.
Its fermionic quasiparticles are correspondingly superposition of electron and hole components. 
The two dimensional, $\mathcal{T}$-symmetry broken topological superconductor is in 
class $\mathbb{D}$ of the Altland-Zirnbauer classification~\cite{Altland1997}. 
The topological phase of this superconductor is identified by an integer $\mathbb{Z}$ index
which counts the number of chiral Majorana edge modes (CMEM). 
When the index $\mathbb{Z}$ is even, the system is topologically equivalent to an ordinary quantum Hall system and 
therefore does not provide a distinct transport signature of chiral Majorana edge modes. 
On the other hand, when the index $\mathbb{Z}$ is odd, the identification of Majorana edge modes and topological superconductivity 
can in principle be accomplished by detecting half quantized Hall plateaus.
This distinction between odd and even-$\mathbb{Z}$, or equivalently between odd and even numbers of 
Majorana edge modes is the starting point for the present work.  

Propagating Majorana modes at the edge of 2D topological superconductors (TSCs) are similar in many ways  
to the Majorana zero modes (MZMs) localized at the ends ~\cite{Alicea2012,Kitaev2001} of 1D TSCs.
MZMs can also be found as zero energy quasiparticles bound to an isolated vortex in a 2D-TSC threaded by 
a unit magnetic flux~\cite{Alicea2012} and are of potential interest as a physical basis for topological quantum 
computation because of their non-Abelian braiding statistics~\cite{Kitaev2003,Nayak2008}. 
A recent proposal also shows that chiral Majorana edge modes can also be used to achieve topological quantum 
computation~\cite{Lian2018a}.

Since natural TSCs are rare, efforts have been made to engineer them. 
The first success was achieved by proximity coupling semiconductor quantum wires with strong Rashba 
spin-orbit coupling to an $s$-wave superconductor under a Zeeman field~\cite{Lutchyn2010,Sau2010,Alicea2010,Alicea2011,Mourik2012,Das2012,Rokhinson2012}.
Topological superconductivity can also be achieved by proximity coupling $s$-wave superconductors to topological insulator surface states ~\cite{Fu2008,Wang2012,Xu2015,Qi2010,He2017,Zeng2018,Chen2018},
placing magnetic ion chains/islands on top of an $s$-wave superconductor~\cite{Nadj2014,Kim2018,Palacio-Morales2018}, 
and occurs in some Fe-based superconductors~\cite{Yin2015,Wang2015a,Wu2016,Zhang2018}. 
For TSC systems based on proximity-coupling to parent $s$-wave superconductors,
the general scheme is to obtain the topological property from an exploitable feature 
of the electronic structure of the host system, for example from the large spin-splitting 
in the Dirac-like surface state of 2D band topological insulators~\cite{Fu2008}. 
Alternately, the topological nature can stem from single particle Chern bands~\cite{Qi2010}.

Experimental work on topological superconductors
has so far focused mainly on finding MZMs at the end of quasi-1D chains~\cite{Mourik2012,Das2012,Rokhinson2012,Xu2015,Nadj2014}. 
Although the Majorana edge modes have been elusive, one recent experiment~\cite{He2017,Qi2010}
has identified possible signatures in a transport study of a magnetic topological insulator thin film that is  
proximity coupled to an $s$-wave superconductor~\cite{He2017}.
In their normal state, magnetic topological insulators can be tuned between 
Chern insulator states that exhibits a quantized anomalous Hall effect which changes sign upon 
magnetization reversal.  In thin films, magnetized topological insulator can also become 
normal insulators.  The Chern insulator is formed when 
interactions between surface state quasiparticles and the magnetic order parameter are dominant and 
the normal insulator state when hybridization between the top and bottom surfaces of the film are dominant. 
When coupled to an $s$-wave superconductor the Chern insulator states yields even-$\mathbb{Z}$ TSCs, 
whereas the normal state generates a topologically trivial superconductor.
If one naively models magnetization reversal by smoothly changing the coupling between quasiparticle 
spins and the magnetic order parameter between positive and negative values, 
odd-$\mathbb{Z}$ superconducting states that support isolated  Majorana edge modes appear
as an intermediate phase.
Because film thickness cannot be tuned {\it in situ}, the transition between normal states with different Chern numbers 
is normally tuned by using an external magnetic field to drive magnetization reversal, generating an intermediate 
state that typically contains a magnetic domain pattern~\cite{Tiwari2017},  complicating the interpretation of 
any experiment~\cite{He2017,Chen2017,Ji2018,Huang2018,Lian2018}.
In this paper, we theoretically explore the possibility of following an alternative and 
potentially simpler route to engineer Majorana edge modes, namely by looking near plateau 
transitions in ordinary quantum Hall systems.  

Since the quantum Hall (QH) effect usually requires fairly 
strong external magnetic fields it has until recently generally been viewed as being incompatible with superconductivity. 
Observing TSC in QH system has recently been identified as an important direction for theoretical and experimental work for 
several reasons: i) to realize parafermions which are generalization of Majorana edge modes obtained when
the fractionalized edge modes of fractional quantum Hall systems have induced topological superconductivity~\cite{Mong2014},
ii) to achieve better tunability and control of multiple Majorana edge modes involving higher Chern numbers, and
iii) because of the on going debate on whether or not a half-integer quantum Hall plateau is a unique 
signature of Majorana edge modes
or can alternately be induced by disorder~\cite{Chen2017,Ji2018,Huang2018,Lian2018}. 
Since QAH edge modes are not typically quantized as perfectly as the QH edge modes in realistic disordered 
systems even in the absence of superconductivity, it seems that progress can be made by looking for half integer quantized Hall 
plateaus in QH systems.  Some theoretical progress in this direction has already been made 
in a study of systems that are proximity coupled to a parent $p$-wave superconductor~\cite{Jeon2019}.

With this motivation, we seek to identify the circumstances necessary to achieve topological superconductivity in the QH regime. 
QH-superconductor proximity 
systems are hybrid-2D systems with the superconductor thin film in
mixed states when under perpendicular magnetic field. 
We show here that vortex lattice properties 
influences the topological properties of QH/$s$-wave hybrid systems.  We attribute this feature to the flatness to single particle Landau 
levels (LL), which lead to bulk quasiparticle dispersion in the superconducting state 
that is largely determined by the vortex lattice periodicity.  
We show that the structure of the vortex lattice plays an important role in Majorana edge mode formation 
by determining the degeneracy of band crossings which can potentially drive topological phase transitions
between states with even and odd-$\mathbb{Z}$ topological indices. 
In particular, we find that the vortex lattices with odd flux per vortex lattice unit cell are equivalent to even parity TSCs and hence only allow even number of chiral Majorana edge modes.
Our findings are most experimentally relevant when a small number of Landau levels host superconducting pairing from an $s$-wave superconductor. 
In this limit, the details of origin of Landau levels whether they arise from Dirac electron or quadratic band are not important for possible topological superconducting phases. 
Instead, the vortex lattice structure plays the main role.

The article is organized as follows. In Sec.~\ref{Sec:model}, we describe a model that includes Landau quantization along with proximity superconductivity. In Sec.~\ref{Sec:VL_states}, we discuss vortex lattice states in 2D and their relationship to electronic
structure in the superconducting state. 
In Sec.~\ref{Sec:bulk}, we discuss topological phases from the point of view of bulk quasiparticle spectra, considering single Landau level limit, then a minimal model containing only two Landau levels, 
and then more generally including many Landau levels with a Debye cutoff.  
For the purpose of illustrating the most important results of this paper, the two Landau level model is the most important case, as well as experimentally achievable.
In Sec.~\ref{Sec:edge} we calculate the edge spectrum and demonstrate 
a transition between ordinary QH edges and edges that support
Majorana edge modes as the bulk topological phase is changed.
In Sec.~\ref{Sec:exp_sys} we discuss possible experiments that are motivated by our 
calculation.  Finally in Sec.~\ref{Sec:discussion}. 
we conclude with a brief discussion of our most important findings.

\section{Pairing in 2D-Dirac Landau levels}\label{Sec:model}

\begin{figure}[!htb]
  \includegraphics[width=0.5\textwidth]{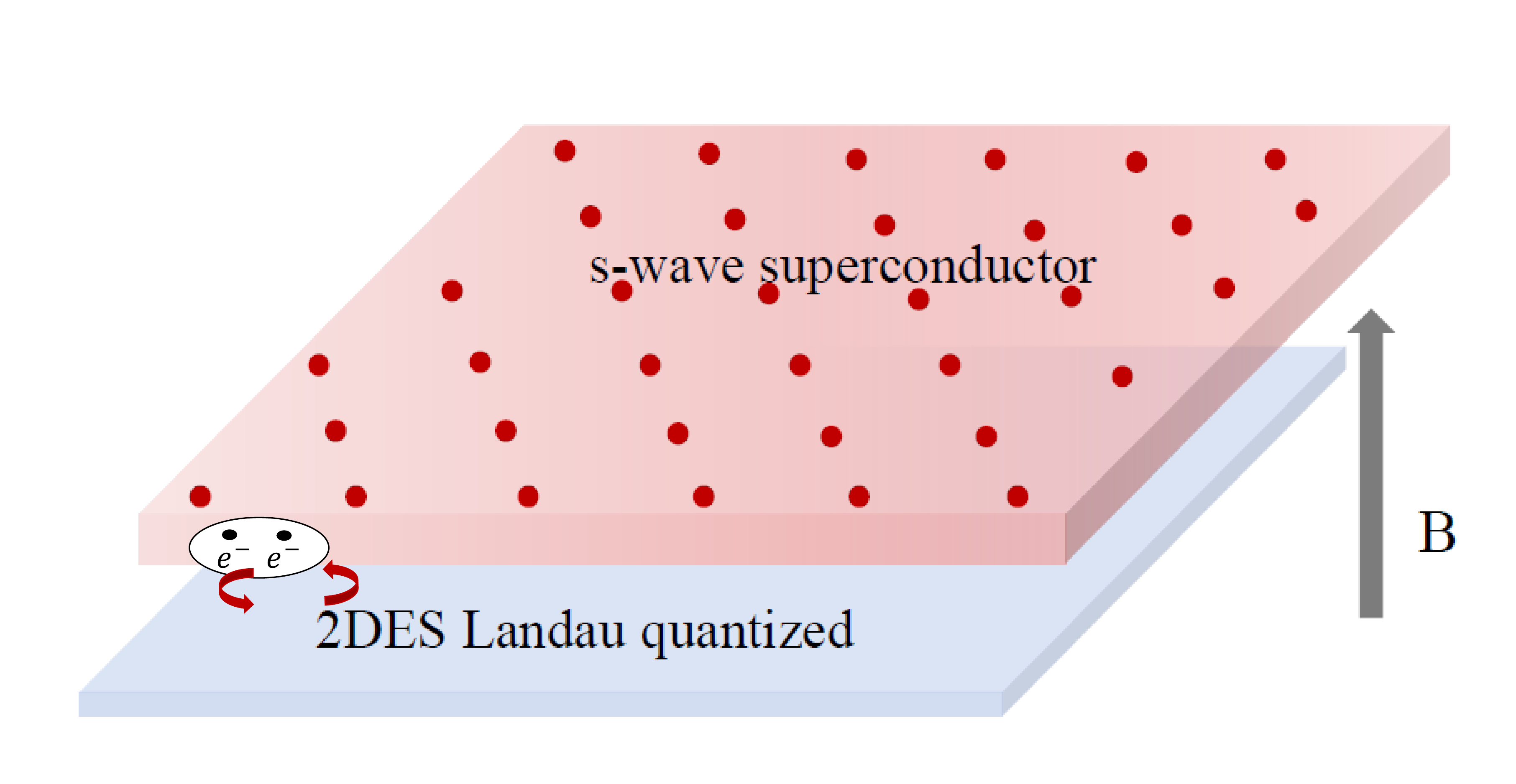}
  \caption{\label{Fig:Setup_QH_TSC}
  Quantum Hall system proximity coupled to an $s$-wave superconductor:
Because of the perpendicular magnetic field, the superconductor exists in the Abrikosov vortex lattice phase. The Cooper pair tunneling back and forth between the superconductor and Landau quantized normal part indicates the proximity effect.}
\end{figure}

In this section we discuss proximity pairing in the Landau levels of a two dimensional system described by a 
linear band crossing (Dirac) model.  
Two dimensional Dirac systems, for example the surface states of three dimensional 
topological insulators, are attractive as hosts for the physics we address because their linear band dispersion leads to 
large Landau level separations and quantum Hall physics at relatively weak magnetic field strengths that are 
more often comparable with superconducting pair potentials.
However, most of our results are general and equally applicable to ordinary 2-dimensional electron gas (2DEG)  systems described by quadratically dispersing bands as we discuss in the 
Appendix~\ref{app:2DEG}. 
We are interested in proximitized superconductivity when these systems are close to 
quantum Hall plateau transitions, {\it i.e.} close to half filling of a Landau level. A schematic of the system under consideration is shown in Fig.~\ref{Fig:Setup_QH_TSC}.  

The BdG Hamiltonian of the proximity coupled system under consideration is
\begin{align}
     & H_{BdG} = \begin{pmatrix}
     \hat{H}_0(\bm{\pi}) & \hat{\Delta}(\bm{r})\\
     \hat{\Delta}^{\dagger}(\bm{r}) & -\hat{H}^{\ast}_0(-\bar{\bm{\pi}})\\
     \end{pmatrix}\, ,
\label{Eq:Single_Dirac_BdG}     
\end{align}
Here, $\hat{H}_0(\bm{\pi}) = -v_F\bm{\pi}\cdot\bm{\sigma}$ is the Dirac Hamiltonian operator of an isolated surface state of a 
three-dimensional topological insulator and $v_F$ is the Dirac velocity.  The Pauli matrices $\sigma_i$ act on electron spin. 
It is important in our calculations that the s-wave  pair potential
\begin{align}
     &\hat{\Delta}(\bm{r}) = \begin{pmatrix}
     0 & \Delta(\bm{r})\\
     -\Delta(\bm{r}) & 0\\
     \end{pmatrix}\, , 
\label{Eq:Pairing_real_spin}     
\end{align}
introduced through proximity coupling is unavoidably non-uniform due to vortex lattice formation
when flux penetrates the adjacent superconductor.
$H_{BdG}$ in Eq.~\ref{Eq:Single_Dirac_BdG} is written in the basis, $\hat{\psi}_{\bm{r}} = (\hat{c}_{\bm{r}\uparrow},\,\hat{c}_{\bm{r}\downarrow},\,\hat{c}^{\dagger}_{\bm{r}\uparrow},\,\hat{c}^{\dagger}_{\bm{r}\downarrow})$, where $\hat{c}_{\bm{r}\uparrow}$ annihilates a spin-up electron at position $\bm{r}$.  We choose to 
introduce the perpendicular magnetic field $\bm{B} = B\hat{z}$ using the Landau gauge 
vector potential $\bm{A} = (-By,\,0)$ which allows for a simple description of the edge modes on 
which we focus much of our attention. The electron and hole subspace kinetic momentum operators are then 
defined as $\bm{\pi} = \bm{p}-eA$ and $\bm{\bar{\pi}} = \bm{p}+eA$ respectively, where 
$\bm{p} = -i\hbar\bm{\nabla}$ is the canonical momentum operator.

\begin{figure*}[!htb]
  \includegraphics[width=1.0\textwidth]{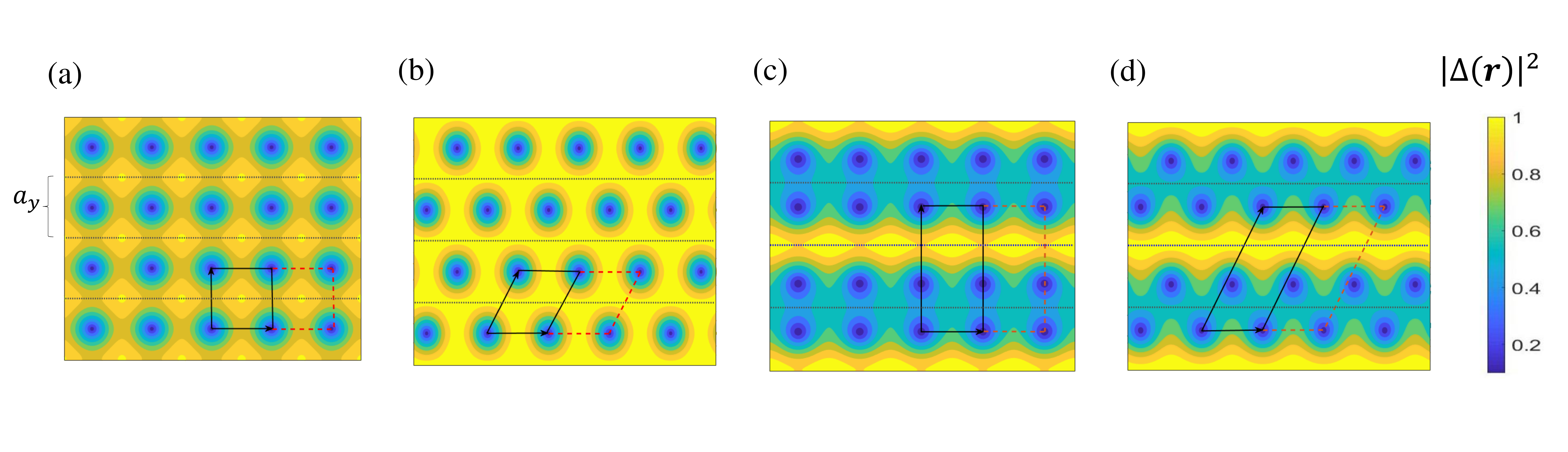}
  \caption{\label{Fig:Vortex_lattice}
  Geometric structure of vortex lattice in two-dimensions. 
  The contributions to the BdG Hamiltonian from adjacent 
guiding centers couple distinct sets of pairs as explained in the main text.
 The solid-black arrows indicate primitive vortex lattice vectors: $\vec{a}_1$ and $\vec{a}_2$.
  The unit cell area they define is indicated by the solid black-lines.  The extended area enclosed by including the dashed-red line is the unit cell in which the BdG Hamiltonian is diagonalized, the dotted horizontal lines indicate the positions
   of the guiding centers $Y = ta_y$ at which the pair amplitude is non-zero. For each vortex of unit vorticity, 
  (a)-(b) two examples of $q=1$  (see main text) odd-flux vortex lattices which have  
  $|\Delta_t| = |\Delta_{t+1}| = \Delta_0$.
  (c)-(d) two examples of $q=2$ even-flux vortex lattices in which $|\Delta_t| \neq |\Delta_{t+1}|$.
    }
\end{figure*}

The BdG Hamiltonian of the two-dimensional Dirac vortex lattice state can be conveniently diagonalized in the Landau level basis.  
We define the Landau level lowering and raising operators in 
the electron subspace of $H_{BdG}$ as 
$\hat{a} = (-\hat{\pi}_x+i\hat{\pi}_y)\ell/\sqrt{2}\hbar$ and $\hat{a}^{\dagger} = (-\hat{\pi}_x-i\pi_{y})\ell/\sqrt{2}\hbar$ respectively. 
In the absence of pairing, Hamiltonian diagonalization reduces to the familiar 
Dirac Landau level problem. The eigenvalues and eigenvectors in absence of pairing are as follows:
\begin{subequations}
\label{Eq:Dirac_LL_eigen} 
\begin{align}
     & \xi_N = \epsilon_0 S_N \sqrt{|N|B}\, ,\\
     &\psi_{N,Y} (\bm{r}) = \mathcal{N}_N\biggl (S_N \phi_{|N|-1,Y}(\bm{r}),\,\,\, \phi_{|N|,Y}(\bm{r}) \biggr )^T\, .     
\end{align}
\end{subequations}
where the normalization factor,
\begin{align}
     & \mathcal{N}_N = 
     \begin{cases}
     \frac{1}{\sqrt{2}}
      & \text{$N \neq 0$}\, ,\\
     1 
     & \text{$N = 0$}\, . 
     \end{cases}
\label{Eq:Dirac_LL_normalization}     
\end{align}
Here, $\epsilon_0 = v_F\sqrt{2e\hbar}$, $S_N$ is the sign of $N^{th}$ Dirac Landau level index, $Y = k_x\ell^2$ is a
guiding center label, $\ell = \sqrt{\hbar/eB}$ is the magnetic length, $\phi_{n,k_x\ell^2}(\bm{r}) = (\textup{e}^{ik_xx}/\sqrt{L_x})\,\varphi_{n}(y/\ell-k_x\ell)$ is a $n^{th}$ Landau level wave function of the non-relativistic 2DEG, and $\varphi_n(y)$ is a one dimensional harmonic oscillator eigenfunction.

The single particle states in the hole block of $H_{BdG}$ can be constructed similarly
using hole space Landau level raising and lowering operators, $\hat{\bar{a}}^{\dagger} = \hat{\bar{\pi}}_x-i\hat{\bar{\pi}}_y$ and $\hat{\bar{a}} = \hat{\bar{\pi}}_x+i\hat{\bar{\pi}}_y$ respectively, giving the following eigenvalues and eigenvectors:
\begin{subequations}
\label{Eq:Dirac_LL_eigen_hole} 
\begin{align}
    & \bar{\xi}_N = -\epsilon_0 S_N \sqrt{|N|B}\, ,\\
    &\bar{\psi}_{N,Y}(\bm{r}) = \mathcal{N}_N\biggl (S_N \bar{\phi}_{|N|-1,Y}(\bm{r}),\,\,\, \bar{\phi}_{|N|,Y}(\bm{r}) \biggr )^T\, ,
\end{align}
\end{subequations}
where $\bar{\phi}_{N,Y} = \phi^{\ast}_{N,-Y}$. For the convenience of calculating matrix element of the pair potential $\Delta(\bm{r})$, the Bogoliubov quasiparticle amplitude can be expanded over the ordinary 2DEG Landau level basis:
\begin{subequations}
\label{Eq:BdG_quasi_particle}
\begin{align}
    & u^{\nu}_{\sigma}(\bm{r}) = \sum_{N,Y} u^{\nu}_{\sigma,N,Y}\phi_{N,Y}(\bm{r})\, ,\\ 
    &v^{\nu}_{\sigma}(\bm{r}) = \sum_{N,Y} v^{\nu}_{\sigma,N,Y}\bar{\phi}_{N,Y}(\bm{r})\, .    
\end{align}
\end{subequations}
Here $u^{\nu}_{\sigma}(\bm{r})$ and $v^{\nu}_{\sigma}(\bm{r})$ are respectively the spin $\sigma$ electron and hole amplitudes of the $\nu$-th BdG quasiparticle. 

The matrix elements of the pair potential in the 2DEG-Landau level basis are given by 
\begin{align}
    &\mathcal{G}^{N,M}_{Y,Y'} =  \int d\bm{r}\, \Delta(\bm{r})\phi^{\ast}_{|N|,Y}(\bm{r})\bar{\phi}_{|M|,Y'}(\bm{r})\,  .
\label{Eq:Pairing_matrix_element_GC}     
\end{align}
To evaluate the above expression, we first define center of mass (COM) and relative degrees of freedom, using the
Landau level lowering operators,
\begin{align}
    & \hat{a}_R  = \frac{\hat{a}_1+\hat{a}_2}{\sqrt{2}}\, ,\quad 
    \hat{a}_r = \frac{\hat{a}_1-\hat{a}_2}{\sqrt{2}}\, ,
\end{align}
where $\hat{a}_1$ and $\hat{a}_2$ are the Landau level lowering operators of individual electrons of the pair.
In the transformed COM and relative coordinates, the 
wavefunctions $\phi^R$ and $\phi^r$ are identical to the single-particle wavefunctions 
except that the characteristic lengths are scaled to account for the changes of charge and mass. 
The effective magnetic lengths are $\ell^R = \ell/\sqrt{2}$ and $\ell^r = \sqrt{2}\ell$, for the COM and relative eigenstates respectively. 
Using this transformation the pairing matrix elements can be evaluated with the result:(See Appendix~\ref{app:Pairing_LL} for a derivation.)

\begin{align}
 &\mathcal{G}^{N,M}_{Y,Y'} 
    =  \sum^{\infty}_{j = 0}\,\mathcal{B}^{N,M}_j\chi_{|N|+|M|-j}(Y_r)F_j\,\delta_{Y'_C,Y_C}\, .
\label{Eq:app_pairing_channel}    
\end{align}
Here $Y_C$ and $Y_r$ are COM and relative guiding centers respectively and the 
$\delta_{Y_C,Y'_C}$ term captures conservation of COM guiding center during a scattering event, and 
\begin{align}
    &\chi_{j}(Y) = \varphi _{j}\biggl (-\frac{Y}{\sqrt{2}\ell}\biggr )\,
\label{Eq:phi_chi}    
\end{align}
is associated with relative Landau level wavefunction.  
The $F_j$ pair-potential strength parameter in Eq.~\ref{Eq:app_pairing_channel} are dependent on the details of the vortex
lattice of the host superconductor and the proximity coupling process, but are expected to be dominated 
by the $j=0$ ( minimum center-of-mass kinetic energy ) term.  
The $\mathcal{B}^{N,M}_j$ define 
the unitary transformation of pair states from single-particle to COM and relative coordinates:
\begin{align}      
    & \mathcal{B}^{N,M}_j = \sum^j_{m=0} (-)^{M-m}\sqrt{\frac{{^j}C_m {^M}C_m {^N}C_{j-m} {^{N+M-j}}C_{M-m}}{2^{N+M}}}  \,  ,
\label{Eq:app_transformation_identitites}
\end{align}
where ${^n}C_k = n!/[k!(n-k)!]$ is a binomial coefficient, $N$ and $M$ are single-particle Landau level indices, 
and $j$, $N+M-j$ are the COM and relative Landau level indices~ \cite{MacDonald1992,Norman1995} 
In vortex lattice states the COM Landau level decomposition is dominated 
by the $j=0$ channel over a broad range of perpendicular magnetic field strengths,
as can be verified using  semi-classical solutions of the non-linear Ginzburg-Landau 
equation~\cite{Tinkham1996,MacDonald1992}.
Truncating the pair Hilbert space to $j = 0$ simplifies the BdG calculations described below,
mainly by reducing the number of parameters needed to specify the pair potential to one strength parameter.

\section{Vortex lattice states\label{Sec:VL_states}}
As mentioned above, we expect that the thin film superconductor 
responsible for proximitized superconductivity  
will form a vortex lattice phase over a broad range of perpendicular magnetic field below $H_{c2}$. 
In Fig.~\ref{Fig:Vortex_lattice}, we show geometrical structure of some typical vortex lattices considered in this paper.
We will show that, if each vortex has unit vorticity,  the geometric vortex lattices shown in Fig.~\ref{Fig:Vortex_lattice} (a)-(b) fall in the same class and do not allow odd $\mathbb{Z}$ TSC phase, while vortex lattices shown in Fig.~\ref{Fig:Vortex_lattice} (c)-(d) can allow odd $\mathbb{Z}$ TSC phase.
As we will explain further below,  the way in which translational symmetry is used to block-diagonalize the BdG Hamiltonian depends on the precise lattice structure.
Assuming that the COM Landau level index is $j=0$, and choosing a Landau gauge with guiding center orbitals extended along one of the primitive lattice vectors of the vortex lattice, it can be shown that the pair amplitude can be non-zero only on a set of equally spaced pair guiding centers: $Y = t a_y$, where $t$ is an integer and $a_y$ is a parameter related to the geometric structure of the vortex lattice and discussed in the next paragraph.
For short-range interactions and vortices of unit vorticity, the resulting pair potential in the Landau gauge has the form~\cite{Norman1995},
\begin{align}
\label{eq:Delta}
    \Delta(\bm{r}) = \sum_t \Delta_t  \, \phi_{0,\sqrt{2}ta_y}(\sqrt{2}\bm{r} )\, .
\end{align}
Here $t \in \mathcal{Z}$ varies over integers and the dependence of 
$\Delta_t$ on $t$ determines the vortex lattice structure.

We define the vortex lattice based on the 
translational symmetries of the magnitude of the pair potential in Eq.~\ref{eq:Delta}. 
For translation $\lambda a_x$ in the $x$-direction where $\lambda$ varies continuously
and $qa_y$ in the $y$-direction where $q\in \mathcal{Z}$,
\begin{align}
    &|\Delta(x+\lambda a_x,y+q a_y)| = [\sum_{t,t'} \Delta^{\ast}_{t'+q}\Delta_{t+q} \textup{e}^{2i(t-t')\lambda \frac{a_xa_y}{\ell^2}}\notag\\
    &\hspace{2cm}\times \varphi_0\biggl (\frac{\sqrt{2}}{\ell}(y-t'a_y)\biggr )\varphi_0\biggl (\frac{\sqrt{2}}{\ell}(y-ta_y)\biggr )]^{1/2}\, .
\label{Eq:Pair_potential_translation1}    
\end{align}
For $\lambda = 1$ and $q = 0$ the translation is by $a_x$ in the  $x$-direction. For unit vorticity, $a_x a_y = \pi\ell^2$, and the magnitude of the pair potential is invariant under this translation, \textit{i.e.} $|\Delta(x+a_x,y)| = |\Delta(x,y)|$. In fact with our gauge choice the pair potential itself is invariant for a pure $x$-translation by $a_x$, without taking the absolute value.
It follows that $\vec{a}_1 = a_x \hat{x}$ is one of the two primitive lattice vectors of the vortex lattice.  
For $q\neq 0$, the magnitude of the pair potential after translation along the $y$-direction is
\begin{align}
    &|\Delta(x+\lambda a_x,y+q a_y)| = [\sum_{t,t'} \Delta^{\ast}_{t'+q}\Delta_{t+q} \textup{e}^{2i(t-t')\lambda \pi }\notag\\
    &\hspace{2cm}\times \varphi_0\biggl (\frac{\sqrt{2}}{\ell}(y-t'a_y)\biggr )\varphi_0\biggl (\frac{\sqrt{2}}{\ell}(y-ta_y)\biggr )]^{1/2}\, .
\label{Eq:Pair_potential_translation2}     
\end{align}
In general translational symmetry in the $y$-direction must be accompanied by translation in the $x$-direction; \textit{i.e.} the 
second vortex lattice primitive vector need not be perpendicular to the first.
Further, for this particular Landau gauge choice, only the absolute value of the pair potential can be made 
invariant under the second translation, while the actual pair potential picks up a phase. 
When $\Delta_t$ is independent of $t$, invariance occurs for $q=1$ and $\lambda=0$ so that 
the second translation vector is $a_y \hat{y}$ and the vortex lattice is rectangular (the special case of $a_x = a_y$ in this class is the square vortex lattice).
We define $q$ as the minimum non-zero integer required to satisfy $|\Delta(x+\lambda a_x, y+q a_y)| = |\Delta(x,y)|$.
For general $\lambda$ and $q$, periodicity is achieved when 
\begin{align}
    &\Delta_{t+q} = \textup{e}^{-i2\pi \lambda t} \textup{e}^{i\theta} \Delta_{t}\, ,
\label{Eq:recursion_VL}    
\end{align}  
is satisfied for some value of $\theta$. 
For example $q = 1$ and $\lambda = 1/2$ yields
\begin{align}
   &\Delta_t = \textup{e}^{-i\pi(t-1)} \textup{e}^{i\theta} \Delta_{t-1}\notag\\
   &\hspace{0.4cm}= \textup{e}^{-i\frac{\pi}{2} t^2}\textup{e}^{i(\frac{\pi}{2}+\theta)t}\Delta_0\, ,
\end{align}
which defines a triangular vortex lattice.  The specific choice of $\theta = -\pi/2$ gives the familiar 
expression $\Delta_t = \exp(-i\pi t^2/2)\Delta_0$ often used for a triangular vortex lattice.
The second primitive lattice vector is then $\vec{a}_2= \lambda a_x\hat{x} + q a_y\hat{y}$.  The vortex lattice unit cell has area 
$\hat{z} \cdot (\vec{a}_1\times \vec{a}_2)$,
and contains $q$ superconducting flux quanta.
Below, we refer to vortex lattices as even or odd, depending on whether $q$ is even or odd.
Theoretically, $q = 1$ odd vortex lattices are the most commonly studied.

In anticipation of the vortex lattice symmetry properties explained in the previous paragraph, 
we partition guiding centers $Y$ into discrete and continuous contributions by writing 
$Y = s a_y+y$ where $s\in\mathcal{Z}$ and $y\in [0,a_y)$. 
Since we are considering pairing in Dirac systems in the quantum Hall limit,  we 
express $H_{BdG}$ first in the Dirac-Landau level guiding center basis.
The BdG equations then take the form
\begin{subequations}
\label{Eq:Eigen_LL_basis}
\begin{align}
     & (\xi_N-\mu) u_{N,s}(y)\, + \, \sum_{M,s'} \mathcal{F}^{N,M}_{s,s'}(y) v_{M,s'} = E u_{N,s}(y)\, ,\\
     &  \sum_{M,s'} (\mathcal{F}^{M,N}_{s',s}(y))^{\ast} u_{M,s'}(y) +  (\mu-\xi_N) v_{N,s}(y) \, \notag\\
     &\hspace{5cm} = E v_{N,s}(y)\, .
\end{align}
\end{subequations}
Here we have transformed from expansion over 2DEG-Landau levels in Eq.~\ref{Eq:BdG_quasi_particle} to the Dirac-Landau levels. Consequently, for $j=0$ pairing we find that (See appendix ~\ref{app:Pairing_LL}, Eq.~\ref{Eq:app_Pairing_matrix_GC_Dirac_step}) 
\begin{align}
    & \mathcal{F}^{N,M}_{s,s'}(y) = \sum_{t} \Delta_{t} \mathcal{D}^{N,M}_0\delta_{s+s', 2t}\notag\\
    &\hspace{2cm}\times\chi_{|N|+|M|-1}((s-s')a_y+2y) \, ,
\label{Eq:Pairing_matrix_element}     
\end{align}
with
\begin{align}
    & \mathcal{D}^{N,M}_0 = S_M \mathcal{B}^{|N|,|M|-1}_0-S_N \mathcal{B}^{|N|-1,|M|}_0\, .
\label{Eq:Unitary_Dirac_LL}    
\end{align}  
Note that $s$-wave pair potentials do not pair electrons that are both in $N=0$ Dirac-Landau levels, which are spin-polarized.
However, an electron in the zeroth Dirac-Landau level ($N = 0$) does pair with electrons from 
higher Dirac-Landau levels ($M\neq 0 $).

Eq.~\ref{Eq:Eigen_LL_basis} organizes the degrees of freedom into guiding center stripes, 
with each stripe labeled by an integer $s$, and
the guiding centers within a stripe labeled by a continuous variable $y$. 
We note in Eq.~\ref{Eq:Pairing_matrix_element}
that pairing is allowed only between stripes that have indices that are both odd or both even
($\delta_{s+s', 2t}$).  
This property implies that odd and even stripe indices decouple in the BdG equation;
{\it i.e.} we can block diagonalize the entire BdG Hamiltonian into two 
susbsystems distinguished by whether the stripe index is even or odd. 
We  prove in the next paragraph that when $|\Delta_t| = |\Delta_{t+q}|$ and $q$ is odd,
the even and odd $s$ subsystem spectra are identical. 
This distinction between vortex lattice classes plays a major role in distinguishing
topological superconducting phases. We will show that only the even flux vortex lattices allow 
odd-$\mathbb{Z}$ chiral topological superconducting phases.

For an odd value of $q$, we can write
$q = 2p+1$, where $p\in \mathcal{Z}$. 
We now examine how the pairing matrix elements in the odd subsystems are related to the pairing matrix 
elements in the even subsystems.  From Eq.~\ref{Eq:recursion_VL} and Eq.~\ref{Eq:Pairing_matrix_element}
\begin{align}
     &\mathcal{F}^{N,M}_{2n+1,2m+1} = \textup{e}^{-i2\pi\lambda(n+m-2p)}\textup{e}^{i\theta}\mathcal{F}^{N,M}_{2(n-p),2(m-p)}\,,
\label{Eq:odd_even_relation}     
\end{align} 
where $n, \, m \in \mathcal{Z}$.  Note that $s = 2n+1$ and $s' = 2m+1$  are odd subsystem stripe indices, and $2(n-p)$ and $2(m-p)$
are even subsystem stripe indices. 
Using Eq.~\ref{Eq:Eigen_LL_basis}, the BdG eigenvalue equations for the odd subsystem can be written as,
\begin{subequations}
\label{Eq:Eigen_LL_basis_odd}
\begin{align}
     & (\xi_N-\mu) u_{N,2n+1}(y)\, +\, \sum_{M,m}[ \mathcal{F}^{N,M}_{2(n-p),2(m-p)}(y)\notag\\
     &\hspace{1cm}\times \textup{e}^{-i2\pi\lambda(n+m-2p)}\textup{e}^{i\theta}v_{M,2m+1}] = E u_{N,2n+1}(y)\, ,\\
     &  \sum_{M,m} [(\mathcal{F}^{M,N}_{2(m-p),2(n-p)}(y))^{\ast} \textup{e}^{i2\pi(n+m-2p)i}\textup{e}^{-i\theta}u_{M,2m+1}(y)]\notag\\
     & \hspace{1cm}+  (\mu-\xi_N) v_{N,2n+1}(y) = E v_{N,2n+1}(y)\, .
\end{align}
\end{subequations}
After the unitary transformation
 $(\bar{u}_{N,2n+1},\, \bar{v}_{N,2n+1})^T = S_n (u_{N,2n+1},\, v_{2n+1})^T$ with 
\begin{align}
     S_n = \begin{pmatrix}
     \textup{e}^{i\theta /2}\textup{e}^{2i\pi\lambda (n-p)} & 0\\
     0 & \textup{e}^{i\theta /2}\textup{e}^{-2i\pi\lambda (n-p)},
     \end{pmatrix}
\end{align}
the BdG eigenvalue equations are transformed to,
\begin{subequations}
\label{Eq:Eigen_LL_basis_odd_even_proof}
\begin{align}
     & (\xi_N-\mu) \bar{u}_{N,2n+1}(y)\notag\\
     &+ \sum_{M,m} \mathcal{F}^{N,M}_{2(n-p),2(m-p)}(y) \bar{v}_{M,2m+1} = E \bar{u}_{N,2n+1}(y)\, ,\\
     &  \sum_{M,m} [(\mathcal{F}^{M,N}_{2(m-p),2(n-p)}(y))^{\ast} \bar{u}_{M,2m+1}(y)]\notag\\
     & \hspace{1cm}+  (\mu-\xi_N) \bar{v}_{N,2n+1}(y) = E \bar{v}_{N,2n+1}(y)\, ,
\end{align}
\end{subequations}
which are identical to those of the even subsystem.
Hence, the odd and even subsystems for an odd-$q$ vortex lattice are equivalent.  
It is easy to see this degeneracy between odd and even subsystem is in general lifted in case of even-$q$ vortex lattices, 
since there is then no similarity relation like Eq.~\ref{Eq:odd_even_relation}.

In the above formulation, we have assumed that each vortex has unit vorticity.
In principle, each geometrical vortex can have higher vorticity of the order parameter phase.
For example, if the each vortex has a vorticity of two then even for the $q = 1$ geometric vortex lattices shown in the Fig.~\ref{Fig:Vortex_lattice}(a)-(b), the unit cell contains even superconducting flux.
In App.~\ref{app:higher_vor}, we demonstrate using the case of vorticity two and three that our rule still follows, \textit{i.e.} vortex lattices with even superconducting flux per unit cell do not allow odd-$\mathbb{Z}$ TSC phases.

In the limit $\Delta(\bm{r})\rightarrow 0$, the system has continuous magnetic translation 
symmetry.  In anticipation of the symmetries of the class of vortex lattice states
that we wish to consider, we can exploit the discrete magnetic translational 
symmetry that remains by choosing unit cells that contain integer numbers of 
{\it electron} magnetic flux quanta. 
To be concrete we choose 
$A_x A_y = 2q\pi\ell^2$, where $A_x = 2a_x$ and $A_y = qa_y$.
With this choice the BdG problem for any general vortex lattice can be block diagonalized in a magnetic Bloch state basis set:
\begin{align}
    &\phi_{N,n,\bm{k}}(\bm{r}) = \sqrt{\frac{qa_y}{L_y}}\sum_t \textup{e}^{ik_y(qt+n)a_y}\textup{e}^{i\pi\lambda q t(t-1)/2}\notag\\
    &\hspace{2cm}\times \textup{e}^{i(\pi\lambda n-\theta/2)t} \phi_{N,k_x\ell^2+(qt+n)a_y}(\bm{r})\, .
\end{align}     
where  integer $n\in [0,..,q-1]$ and $\bm{k} = (k_x,\, k_y)$ is a Bloch wave vector with 
$k_x\in [0,\,\pi/a_x)$ and $k_y\in [0,\,2\pi/qa_y)$.
The pairing in magnetic Bloch basis is a $q\times q$ matrix for every pair of Landau level indices $N, M$ of pairing electrons:
\begin{align}
     &\mathcal{F}^{N,M}_{n,m}(\bm{k}) = \mathcal{D}^{N,M}_0\sum_{t}\Delta_{qt+(m+n)/2}\,\textup{e}^{-i\pi \lambda q t (t-1)}\notag\\
     &\hspace{2cm}\times\textup{e}^{-i(2\pi\lambda n -\theta)t}\textup{e}^{ik_ya_y(2qt+n-m)}\notag\\
     &\hspace{2cm}\times\chi_{|N|+|M|-1}((2qt+n-m)a_y+2k_x\ell^2)\, .
\label{Eq:Pairing_matrix_k}     
\end{align}
Since $\Delta_t$ is only non-zero on integer values of $t$, the pairing matrix element in $\bm{k}$-space is only non-zero when $n$ and $m$ are either both even or both odd.
Based on the arguments used to prove the degeneracy for odd-flux vortex lattice, \textit{i.e.} $q = 2p+1$ from the real space picture, the degeneracy can also be proven in the $\bm{k}$-space picture here. 
More precisely, in the even-odd block diagonalized system,  the odd system with $n = 2s+1$ and $m = 2s'+1$ at $\bm{k} = (k_x,\, k_y)$ is degenerate with the even system with $n = 2(s-p)$ and $m = 2(s'-p)$ at $\bm{k} =  (k_x, k_y+\frac{\pi}{2q a_y} \,)$. The algebra and linear transformation to show that follows exactly like Eq.\ref{Eq:odd_even_relation}-\ref{Eq:Eigen_LL_basis_odd_even_proof}.

So far we have formally proven that for the system in consideration here, 2D vortex lattices have distinct classification into odd-flux and even-flux vortex lattice, where only the even-flux vortex lattice can host odd-$\mathbb{Z}$ TSC phases.
The equivalent physical statement is that the odd flux vortex lattice leads to an even parity topological superconductor.
In the strong magnetic field case, however, the analog of a Kramers pair occurs in our gauge choice between an electron at $(k_x, k_y)$ with an electron at $(k_x, k_y+\frac{\pi}{2q a_y} \,)$ instead of conventional $\bm{k},-\bm{k}$ Kramer pairs.

To simplify the further discussion related to degeneracy in the spectrum, first note that irrespective of type of vortex lattice, the pairing matrix element in Eq.~\ref{Eq:Pairing_matrix_k} has an important translational property:
\begin{align}
    & \mathcal{F}^{N,M}_{n,m}(k_x,k_y+\pi/a_y) =\mathcal{F}^{N,M}_{n,m}(k_x,k_y)\, .
\end{align}
For $q=1$ vortex lattices the period is half the reciprocal lattice vector. This implies
that the BdG spectrum at a point $\bm{k_0} = (k_{x0},\, k_{y0})$ is identical to the spectrum at $\bm{k_1} = (k_{x0},\, k_{y0}+\pi/a_y)$. For the simplest even flux vortex lattices (\textit{i.e.} $q = 2$), the reciprocal space lattice vector length along 
$k_y$ is equal to $\pi/a_y$, and this degeneracy is not present.

\section{Bulk picture}\label{Sec:bulk}
Having established all the necessary framework and formal proofs related to distinction between the BdG spectral degeneracies of 
even and odd flux vortex lattices, we now discuss how spectra depend on
the number of Landau levels contained within a pairing window. 
We will discuss the cases of pairing within one Landau level, two Landau levels,
and many Landau levels separately.
Since our main goal is to address the connection between vortex lattice structures and the topological phases, 
we study $q = 2$ vortex lattices as the simplest example of the even
 flux vortex lattices and compare with $q = 1$ odd-flux vortex lattices.  
In addition, we restrict our attention to the $\lambda = 0$ and $\lambda = 1/2$ cases, 
which for $q = 1$ give square and triangular vortex lattices respectively. 
For the most part, we will be solving the BdG matrix equation:
\begin{subequations}
\label{Eq:BdG_diag_k_space}
\begin{align}
     &(\xi_N-\mu)u_{N,n}(\bm{k})+\sum_{M,m}\mathcal{F}^{N,M}_{n,m}(\bm{k})v_{M,m}(\bm{k})\notag\\
     &\hspace{3cm} = E(\bm{k})u_{N,n}(\bm{k})\, ,\\
     & \sum_{M,m}(\mathcal{F}^{N,M}_{n,m}(\bm{k}))^{\ast}u_{M,m}(\bm{k})+(\mu-\xi_N)v_{N,n}(\bm{k})\notag\\
     &\hspace{3cm} = E(\bm{k})v_{M,m}(\bm{k})\, ,
\end{align}
\end{subequations}
for $q = 1$ and $q = 2$. For the case of $q = 1$ we will often not use subscript $n,m$ indices.
\begin{figure}[!htb]
  \includegraphics[width=0.5\textwidth]{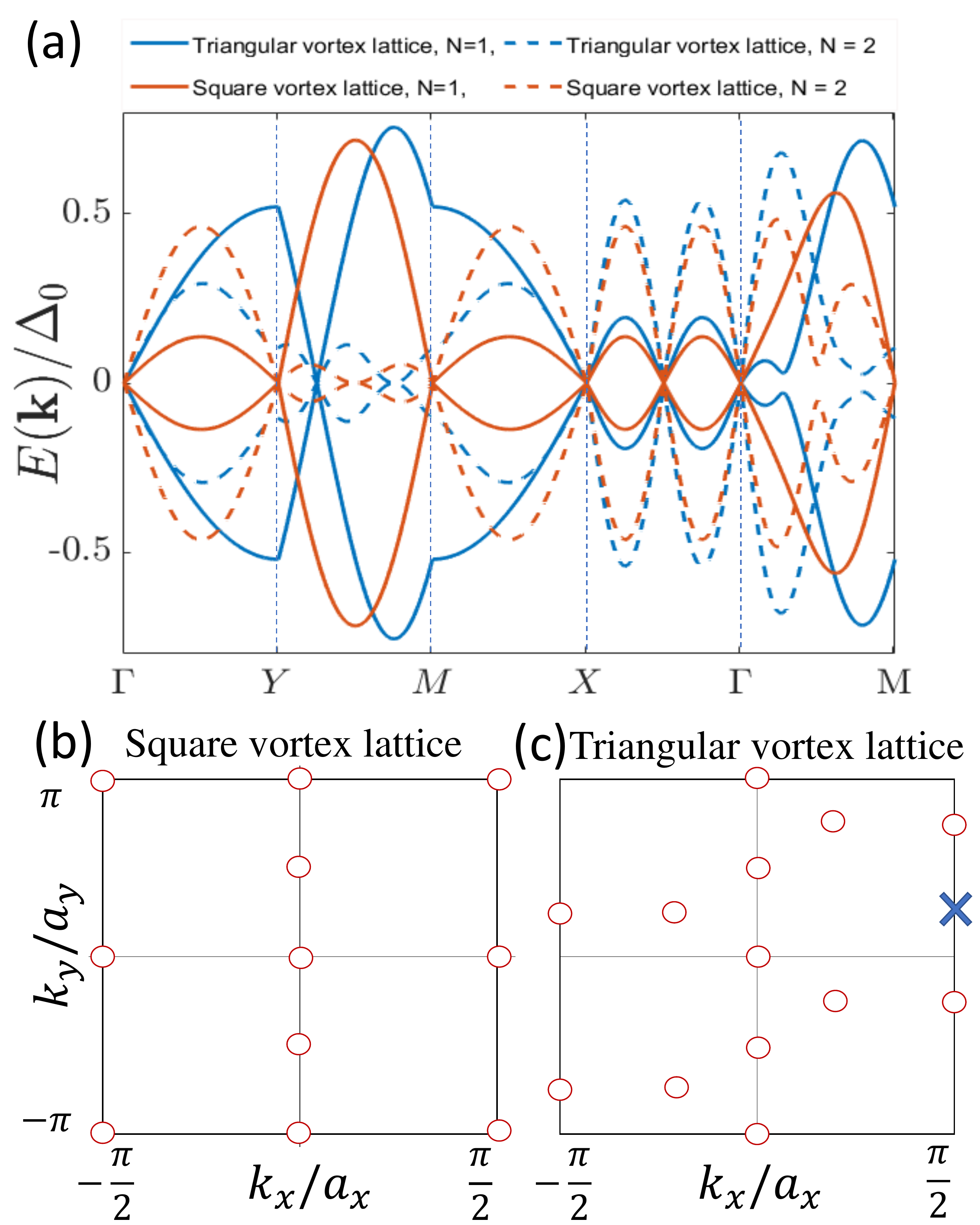}
  \caption{\label{Fig:Single_LL_spectrum}
  (a) Bogoliubov quasiparticle spectrum with the BdG operator projected onto the 
  $N=1$ and $N=2$ Landau levels for triangular (blue) and square (red) vortex lattices. 
  (b)-(c) Band- touching Dirac point positions indicated by the red circles for pairing in $N = 1$ Dirac Landau levels 
  for square and triangular vortex lattices. 
   The band touching points at $(k_{x0},\, k_{y0})$ are equivalent to
   those at $(k_{x0},k_{y0}+\pi/a_y)$.
   Notice that triangular vortex lattice case respects the $C_6$ rotation symmetry with respect to the point indicated by blue cross as the center of rotation. The $k_y$ axis scales twice of $k_x$ axis in (b) and (c).
  }
\end{figure}

\subsection{Pairing within a single Landau level}

The simplest example of superconductivity in the QH regime is the case in which only one 
Landau level lies within the pairing window.  This limit is relevant if a regime can be achieved in which 
the Landau level separation is larger than the Debye pairing window energy $E_D$ of the parent superconductor and
at the same time magnetic field weaker than its upper critical field $H_{c2}$.  This is, of course, not a regime that 
is frequently achieved, but is more accessible when the proximitized system has a Dirac spectrum with widely spaced 
low energy Landau levels that a parabolic system with equally spaced Landau levels.  
Because of Landau level truncation, the BdG spectrum in this limit is simply given by
\begin{align}
     & E(\bm{k}) = \pm\sqrt{(\xi_N-\mu)^2+|\mathcal{F}^{N,N}(\bm{k})|^2}\, .
\label{Eq:Single_LL_spectrum}     
\end{align}
When the relevant Landau level energies are aligned with the chemical potential, quasiparticle energies vanish 
whenever $\mathcal{F}^{N,N}(\bm{k}) = 0$. These positions in momentum space at which the energies 
vanish are then related to the zeros of the Hermite polynomials associated with the pairing Landau level, 
as noted previously~\cite{Zocher2016} in relation to spinless pairing in the zeroth-Landau level. 
For the simplest case of pairing in the $N = 1$ Dirac Landau level, all the band touching points are Dirac like as seen in Fig.~\ref{Fig:Single_LL_spectrum}.  
If pairing is in $N>1$ Dirac Landau level,   
as illustrated in Fig.~\ref{Fig:Single_LL_spectrum}(a), both linear and quadratic band touching points occur.  
For odd $q$ flux lattices each band touching point that appears in the interval $[-\pi/qa,0)$ are replicated in the interval $[0,\pi/qa)$, as explained in the previous section.  

To further explore the topological nature of the physics in this regime we first assume $\mu$ to be slightly away from $\xi_N$, so that spectrum becomes gapped. Next, we write an effective $2\times 2$ low energy effective Hamiltonian describing the system near the previously (when $\xi_N \approx \mu$ )band touching points,
\begin{align}
     H^{\textup{eff}}_{BdG}\sim \begin{pmatrix}
     -\mu & (\alpha k_y\pm i\beta k_x)^{\gamma}\\
     (\alpha k_y\mp i\beta k_x)^{\gamma} & \mu
     \end{pmatrix}\, .
\label{Eq:Single_LL_Low_E}     
\end{align}
Here $k_x, k_y$ are measured relative to some point $\bm{k}_0$ in BZ, where $\mathcal{F}^{N,N}(\bm{k}_0) = 0$, we have chosen the zero of energy to be the single-particle energy of the relevant Landau level and the $\pm$ sign allows either chirality for the Dirac points, and the power $\gamma$ is lowest order for which
\begin{align}
    & \frac{\partial^{\gamma}\mathcal{F}^{N,N}(\bm{k})}{\partial k^{\gamma}_{x/y}}\biggr |_{\bm{k} = \bm{k}_0} \neq 0\, ,
\end{align}
for example  $\gamma = 1$ for Dirac band touching and $\gamma = 2$ for quadratic band touching in the limit $\mu = \xi_N$. 
In Eq.~\ref{Eq:Single_LL_Low_E}
and $\alpha$ and $\beta$ are constants that depend on the Fermi velocity of the low energy Bogoliubov quasiparticle at $\bm{k}_0$. 
The use of different $\alpha$ and $\beta$ allows for the anisotropies in band touching point velocities that 
can be seen in Fig.~\ref{Fig:Single_LL_spectrum}(a).  
The Landau level is partially occupied when $|\mu| \lesssim \Delta_0$.
For Dirac band touching, the low energy $H^{\textup{eff}}_{BdG}$ in Eq.~\ref{Eq:Single_LL_Low_E} resembles the BdG matrix of a 
spinless $p$-wave superconductor near the critical point of topological phase transition. 
The two topologically distinct phases are distinguished by the sign of $\mu$.
When $\mu$ is tuned through zero, gaps close and reopen and the system experiences 
topological phase transitions at which the total Chern number changes.
In finite systems, the number of edge state channels also changes, as we discuss later.
For odd $q$, Dirac points appear in pairs and both Chern numbers, and as we show explicitly later,
the numbers of edge channels change by an even integers when these Dirac points appear.
The quadratic band touching points relevant to pairing in 
higher Landau levels can be analyzed in a similar way.

\begin{figure*}[!htb]
  \includegraphics[width=1.0\textwidth]{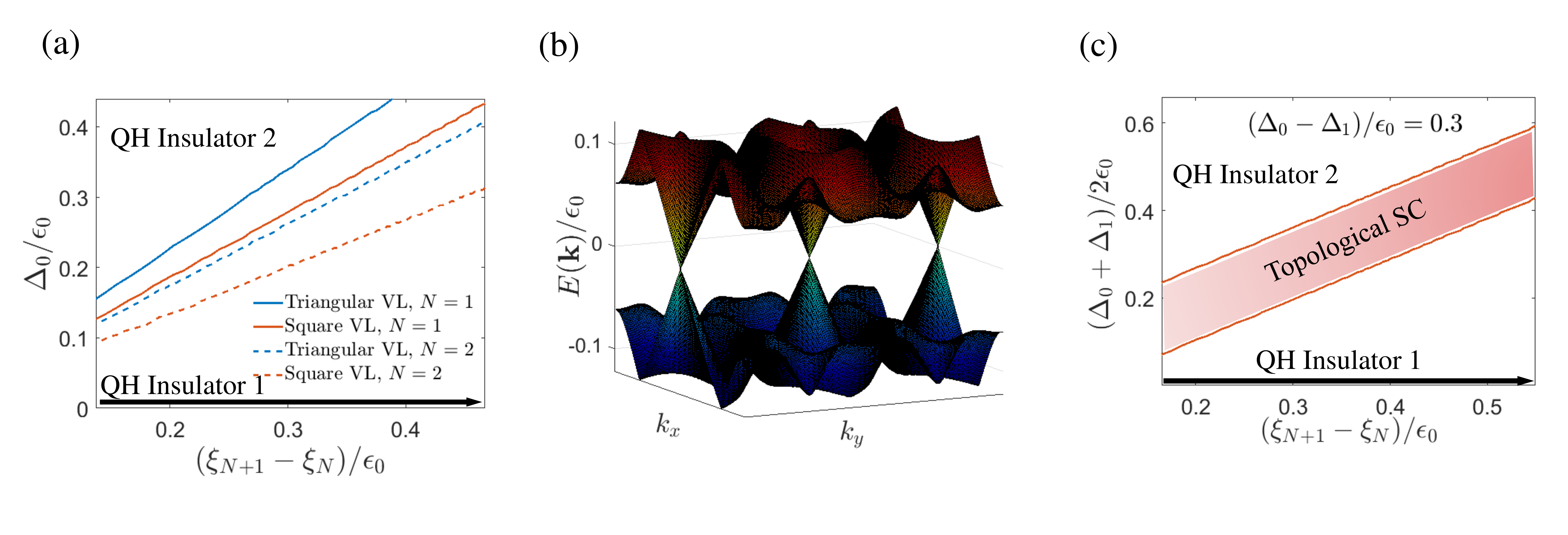}
  \caption{\label{Fig:2_LL_gap}
  Model with two Dirac Landau levels-Energies are shown in the $\epsilon_0 = v_F\sqrt{2e\hbar}$ units and $\xi_N$ is $N^{th}$ Landau level energy.
    (a) Superconducting pairing strength $\Delta_0$ that leads to the first gap closing as a function of Landau level separation 
    for two Landau level models of the $q=1$ vortex lattices in Fig.~\ref{Fig:Vortex_lattice} (a)-(b).  In these calculations the $\mu$ has been placed halfway between Landau levels so that the lower Landau level is completely occupied and the higher Landau level empty in the absence of pairing.
    Band touchings are accompanied by transitions between a QH insulator (QHI 1) and another QH 
    insulator (QHI 2) state, both of which have an even number of chiral Bogoliubov edge states.
    (b) Quasi-particle spectrum at a gap closing point for $q = 1$ square vortex lattice. 
    In this case there are two Dirac like gap closing points in the first BZ. 
    (c) When the vortex lattice symmetry is lowered by setting the pairing strength $\Delta_t$ to different values 
    in odd and even stripe regions, giving  the vortex lattice illustrated in Fig.~\ref{Fig:Vortex_lattice} (c)-(d), 
    an intermediate chiral topological superconductor (CTSC -red region) phase appears with an odd number of 
    chiral Majorana edge modes.
    }
\end{figure*}

\subsection{Two Landau-level model}\label{Subsec:2_LL_bulk}
In the previous section we discussed topological phase transitions driven by varying 
$\mu$ in a system with a single Landau level in the pairing window.  
In real physical systems it is carrier density that is controlled by gate voltages, not $\mu$, and $\mu$ changes 
irregularly with magnetic field strength.  For example at $T=0$ and in the absence of pairing, 
$\mu$ changes discontinuously, jumping from being pinned at one Landau level energy to begin pinned at 
another.  The minimal model that incorporates that consequence of the strongly peaked densities of states of 
Landau level systems is a model with two Landau levels in the pairing window.  A $q=1$ two Landau level 
system has a $4\times 4$ BdG Hamiltonian which makes it possible to obtain closed form expressions for 
eigenvalues, which are however not especially transparent.  
However, the important insight is that by tuning $\Delta$ or Landau level gap one can close and open BdG gaps.

In the absence of superconductivity, the system is always a gapped quantum Hall insulator (QH 1). 
Gap closings and topological phase transitions can be driven either by varying the pairing strength $\Delta_0$, or by 
using magnetic fields to tune the energy separation between Landau levels, as shown in 
Fig.\ref{Fig:2_LL_gap} (a). The value of $\Delta_0$ at which gap closes, scales with the Landau level separation. 
The circumstance is closely analogous to the case of proximity superconductivity induced in the surface 
states of magnetically doped topological insulators, the system that hosts the only
established experimental example of a quantum anomalous Hall (QAH) state. In that case an intermediate TSC phase appears when 
the surface state exchange fields are reversed to drive the system between two different quantum Hall insulators~ \cite{Qi2010}. 
Based on this analogy one would expect that the first gap closing to 
occur as $\Delta_0$ increases to convert the quantum Hall insulator into a chiral topological superconductor
odd-$\mathbb{Z} = 1$.   Here though the topological phase diagram 
depends on the type of vortex lattice.  In Fig.~\ref{Fig:2_LL_gap} (a) the gap closing lines 
mark phase transitions between QH Insulators. 
The $\Delta_0\rightarrow 0$ limit in this case is a QH state with a full $N=1$ Landau level, 
and as we discuss later in Sec.~\ref{Sec:edge}, two chiral edge states in the doubled BdG Hilbert space.
Once the pairing is turned on the system no longer has quantized Hall conductance, however, it is still adiabatically connected to a QH insulator and has two edge channels as long as the gap does not close. 
The difference compared to the QAH case appears when the gap closes and reopens. 
In the QAH model the Dirac-like gap closings occur only at the $\Gamma$-point in momentum space and 
are generically accompanied by odd integer changes in the topological $\mathbb{Z}$-index.
One of the two BdG doubled quantum Hall edge states survives.
This is the single chiral Majorana edge mode, and is expected to yield a half quantized conductance plateau in transport experiments~\cite{Chung2011}, although the 
experimental reversal process~\cite{He2017} is more complex than simply smoothly changing the surface state exchange field.
When solved in a Landau-level basis, the bulk properties of the QH model do not explicitly 
exhibit the topological character of the non-paired states.   
The BdG spectrum is determined by $\mathcal{F}^{N,M}(\bm{k})$. 
As mentioned earlier, for odd flux vortex lattice,  each energy has even number of $\bm{k}$s. 
Band crossings therefore always occur in pairs as illustrated in Fig.~\ref{Fig:2_LL_gap} (b). 
This leads to the important conclusion that in the QH transition for odd-flux vortex lattices,
the $\mathbb{Z}$-index always changes by an even integer, and the number of edge channels changes by two 
or multiples of two. This conclusion will be confirmed using 
explicit edge state calculations in Sec.~\ref{Sec:edge}. 

To achieve a chiral topological superconducting phase with 
an odd number of edge modes, one needs to break this degeneracy between the even and odd $s$ 
subsystems. The simplest way is to allow different pairing amplitudes, $\Delta_0$ and $\Delta_1$, for even and odd 
index stripes.  With this choice
\begin{align}
    & \mathcal{E}^{N,M}_{s,s'}(y) = \Delta_{0} \textup{e}^{i\theta (s+s')^2} \mathcal{D}^{N,M}_0\notag\\
    &\hspace{2cm}\times\chi_{|N|+|M|-1}((s-s')a_y+2y) \, ,\notag\\
    & \mathcal{O}^{N,M}_{s,s'}(y) = \Delta_{1} \textup{e}^{i\theta (s+s')^2} \mathcal{D}^{N,M}_0\notag\\
    &\hspace{2cm}\times\chi_{|N|+|M|-1}((s-s')a_y+2y) \, .
\label{Eq:Pairing_matrix_element_OE}     
\end{align}
The $\bm{k}$-space picture is modified by block diagonalizing the BdG matrix Eq.~\ref{Eq:BdG_diag_k_space} in the odd and even system and halving the BZ along $k_y$ such that $k_y\in [0,\pi/a_y)$, since this lowers the translation symmetry, such that the smallest repeating unit cell now contains two electronic flux quanta. The matrix takes the following form,
\begin{align}
    &M_{BdG}(\bm{k}) = \begin{pmatrix}
    M^o_{BdG}(\bm{k}) & 0\\
    0 & M^e_{BdG}(\bm{k})
    \end{pmatrix}\, .
\label{Eq:BdG_matrix_halved}    
\end{align}
A change in topological index from even to odd cases occurs when a gap closes
in only one of the two blocks.   Once the degeneracy between odd and even system is broken, the gap closings generically
occur at different values of magnetic field in the even $s$ and odd $s$ blocks.  
The region between the subsystem gap closings is shaded in red in Fig.~\ref{Fig:2_LL_gap} (c). 
We identify this region as having a chiral topological superconductor state and 
an odd number of Majorana edge modes.

\subsection{General model with Debye cut-off}\label{Subsec:Debye_bulk}

\begin{figure}[!htb]
  \includegraphics[width=0.5\textwidth]{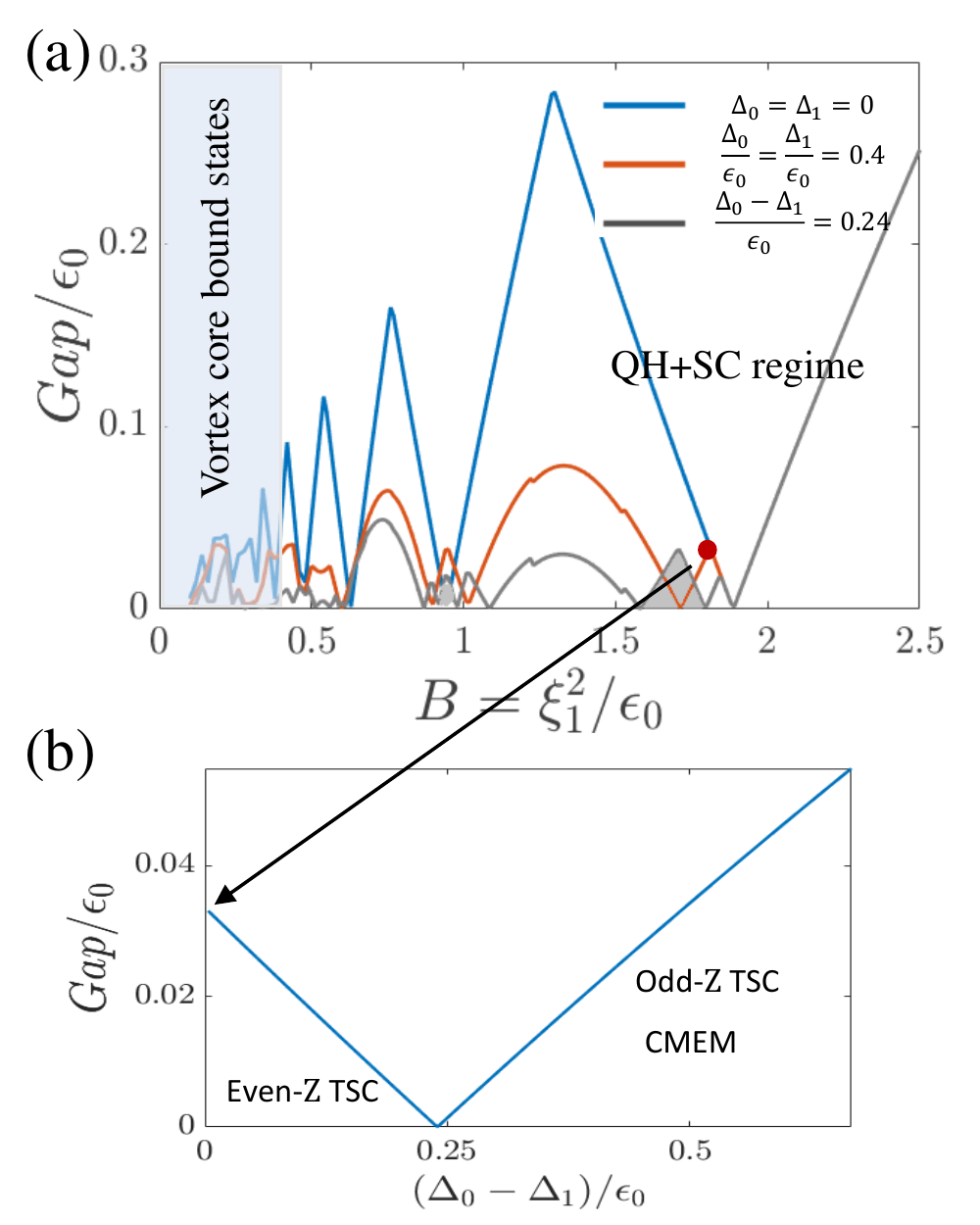}
  \caption{\label{Fig:Gap_Debye}
Bogoliubov quasiparticle gaps (in the $\epsilon_0 = v_F\sqrt{2 e \hbar}$ units), 
(a) as a function of magnetic field strength. Here $\mu = 1.35\epsilon_0$ and a hard Debye cutoff window of $E_D = 1.1\epsilon_0$ is fixed. As the magnetic field is varied the Landau levels which host pairing change.  
The blue curve shows Landau level energies relative to the Fermi level 
in the limit of no pairing. The gap vanishes when the relevant Landau 
level crosses the Fermi energy. 
The plot shows Dirac Landaus with decreasing indices crossing the Fermi level, reaching index $N=1$ at the last 
zero of the blue curve.  The red curve plots gaps as a function of magnetic field 
strength in a $q=1$ vortex-lattice, {\i.e} for a case with 
equal pairing strength in even $s$ and odd $s$ channels.  
Extra gap closings occur as paired Landau levels cross through the Fermi level,
but occur in pairs so that the transitions are between one QH insulator and another.
The grey line calculations is for an $q=2$ vortex lattice in which the pairing strength is different 
for even and odd values of $s$.  Even $q$ allows a chiral topological superconductor 
phase to emerge in the regions that are shaded gray. 
In the shaded blue region towards on left hand side of the figure, the Landau level structure 
is destroyed by pairing and the low energy states are most simply viewed as 
hybridized vortex core bound states.
(b) Gap as a moving from odd-flux to even-flux vortex lattice by tuning $\Delta_0-\Delta_1$. Starting from an even-$\mathbb{Z}$ phase shown by red dot in (a), as $\Delta_0-\Delta_1$ is tuned keeping everything else fixed, extra gap closing occurs, beyond which, one reaches odd-$\mathbb{Z}$ phase. The odd-$\mathbb{Z}$ phase can be further broadened with tuning $\Delta_0-\Delta_1$. 
  }
\end{figure}

In a realistic model the number of Landau levels in the pairing window increases with decreasing magnetic field strength.
In the normal state the number of Landau level below the Fermi level is inversely proportional to magnetic field and 
successive crossings between the Landau levels and the Fermi energy leads to gap 
changes and jumps in Hall conductance.  Here we show that for finite pairing there are extra gap closing points of the BdG spectrum (Gap = $\textup{min}[|E(\bm{k}|]$)  
associated with each Landau level crossing, implying extra topological phase transition points. 
However, for the case of odd-$q$ vortex lattices, all these phase transitions involve simultaneous 
gap closing at two different points in momentum space and connect one quantum Hall insulator 
state with another.  Once the pairing amplitude is allowed to take different values in even and odd striped regions, yielding a 
$q=2$ vortex lattice, band crossings occur singly.  In the example illustrated in Fig.~\ref{Fig:Gap_Debye}
the gray regions show the fields strengths where the ground state is a chiral topological superconductor.
The extent of the CTSC phases can be tuned by varying $\Delta_0-\Delta_1$.  
At weak magnetic fields, the Landau level gap is much smaller than $\Delta_0$, there are many Landau levels within the 
Debye pairing window, and the low energy quasiparticles are best viewed as bands formed by 
hybridizing vortex core bound states associated with different vortices. 
In case of Dirac model, these vortex core bound states at weak magnetic field are MZMs\cite{Liu2015,Murray2015}, since the  Dirac model proximity coupled to an $s$-wave superconductor under time reversal symmetry is the famous Fu-Kane model~\cite{Fu2008}. 
For the 2DEG case at weak field limit, the low energy vortex core bound states are not Majorana. 
In that sense, the low field sector indicated by the 
shaded blue region in Fig.~\ref{Fig:Gap_Debye} is dramatically different for Dirac model and ordinary 2DEG model when proximity coupled to an $s$-wave superconductor, but is not the focus of this work (See Appendix~\ref{app:2DEG} for comparison between Dirac and ordinary 2DEG). However, the high field sector in QH regime is qualitatively same.

\section{Edge state picture}\label{Sec:edge}
\begin{figure*}[!htb]
  \includegraphics[width=1.0\textwidth]{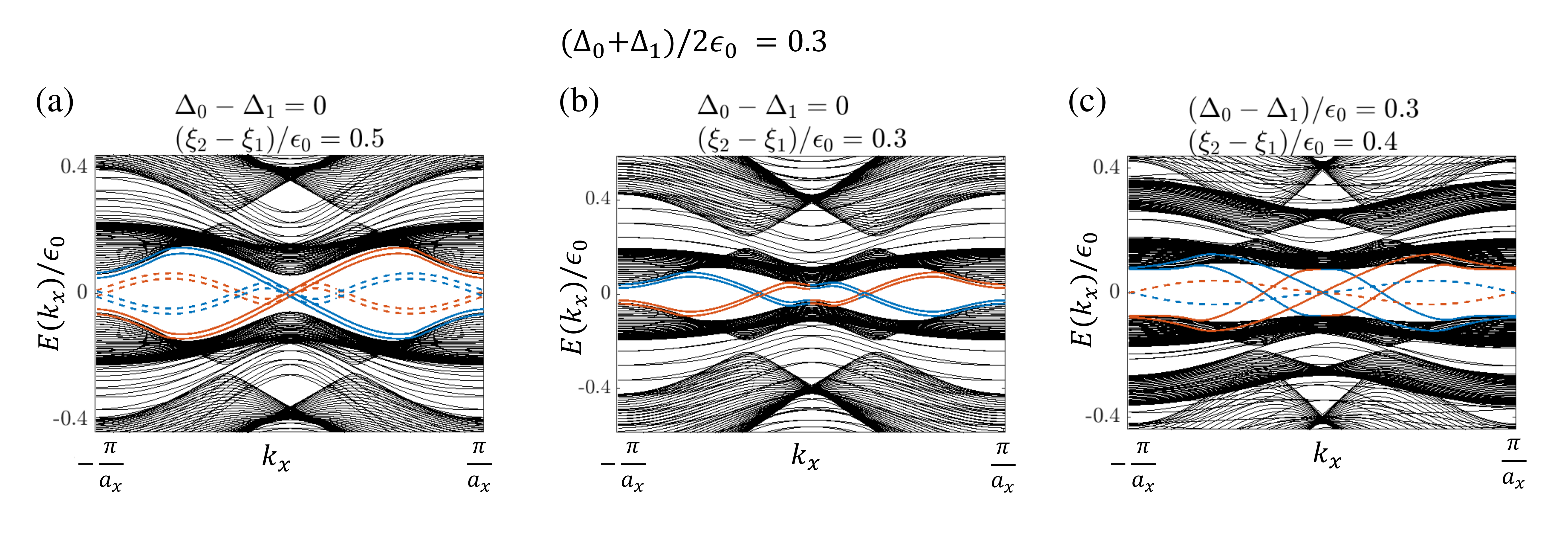}
  \caption{\label{Fig:Edge_states}
  Stripe geometry spectrum (in $\epsilon_0 = v_F\sqrt{2e\hbar}$ units) showing edge states in the gaps between Landau levels. 
  States localized at opposite edges are distinguished by color (blue and red).
  Note that some edge states appear that are not related to the bulk topology.  These are not present 
  at all energies, do not flow between bulk bands, and are distinguished by drawing them with dashed lines. 
  (a) Triangular $q=1$ vortex lattice.  At this magnetic field strength the system has two chiral modes on each edge, 
  corresponding to BdG doubling of the $N=1$ quantum Hall effect.
  (b) Triangular  $q=1$ vortex lattice at a weaker magnetic field.  The system now has four chiral modes on each edge,
  corresponding to a BdG doubling of the $N=2$ quantum Hall effect.  
  (c) For a $q=2$ vortex lattice, ($\Delta_0 \neq \Delta_1$) there is intermediate gap closing point as magnetic field is varied 
  which opens up an interval of field over which the system hosts three chiral modes on each edge. 
  This phase is topologically connected to chiral topological superconductivity.
   }
\end{figure*}

In this section we discuss the BdG spectrum of finite width stripes to establish bulk-edge 
correspondence and demonstrate the presence of chiral Majorana edge modes.
In the continuum Dirac model we employ, we have defined the stripes by 
adding a smooth confining potential around the edge of the sample and truncating the Hilbert space to the 
two Dirac Landau levels. 
In doing so, we have assumed that edge states from Landau levels which are not in the pairing window but 
may be active at the Fermi level do not play a role.  The BdG equations then take the following form:
\begin{subequations}
\begin{align}
     & [\xi_N-\mu+U_s(k_x)]\, u_{N,s}(k_x)\, + \, \sum_{M,s'} \mathcal{F}^{N,M}_{s,s'}(k_x) v_{M,\bm{Y'}}\notag\\
     &\hspace{5cm} = E u_{N,s}(k_x)\, ,\\
     & -[\xi_N-\mu-U_s(-k_x)]\, v_{N,s'}(y)\,\notag\\ 
     &\hspace{5mm}+ \, \sum_{M,\bm{Y'}} (\mathcal{F}^{N,M}_{s,s'}(k_x))^{\ast} u_{M,s'}(k_x) = E v_{N,s'}(k_x)\, .
\end{align}
\label{Eq:Eigen_LL_basis_stripe}
\end{subequations}
Here, the smooth confining is specified by letting 
\begin{align}
     & U_s(k_x) = \begin{cases}
     0 & \text{$ |s| \leq S_{bulk}$}\, ,\\
     U_0(sa_y+k_x\ell^2) & \text{$ |s| > S_{bulk} $}\, .
     \end{cases}
\end{align}
Here $U_0$ sets the strength of confining potential and $S_{bulk}\in \mathcal{Z}$ sets the width of the bulk part of the stripe. 

We have performed stripe state calculations that correspond to the bulk calculations for the 
model that retains only the $N=1$ and $N=2$ Landau levels.  The edge states shown in Fig.~\ref{Fig:Edge_states} accurately 
describe the system 
when $\Delta_0\not\gg \xi_{N+1}-\xi_N$ and the two Landau level model remains valid. Our calculations contain $100$ integer $s$ indices.
In the absence of pairing the system is a QH insulator, 
with edge states at the Fermi level coming from the occupied 
Landau levels.  In the BdG-doubled Hilbert space, these modes are doubled but 
retain the same chirality due to the combination of particle-hole inversion and 
momentum label reversal in the hole block of the BdG equations.  
For weak pairing regime the quantum Hall gaps remain open.
The two chiral edge state branches plotted as solid red and blue lines in Fig.~\ref{Fig:Edge_states}(a), 
are localized on the left and right edges respectively, and evolve from $N=1$ Landau level edge states.  
For the odd $q$ vortex lattices, the bulk calculations discussed in Sec.~\ref{Subsec:2_LL_bulk} and Fig.~\ref{Fig:2_LL_gap} (a) show 
a topological phase transition.  On the other side of the gap closing point the system again has even number of chiral edge states as shown by four chiral edge states in Fig.~\ref{Fig:Edge_states}(b). This evolution is similar to ordinary quantum Hall edge state evolution, as the magnetic field is decreased, the number of edge states increase, and for finite stripe widths are eventually
become indistinguishable from bulk states as $\bm{B}\rightarrow 0$. 
To induce a topological superconducting phase with odd number of Majorana edge mode, we allow even and odd stripes to take different pairing amplitudes, which breaks the degeneracy between the two subsystems. As the magnetic field
strength is decreased, the even and odd subsystem gap closing points move away from each other and the 
intermediate phase is a topological superconductor that hosts three Majorana edge modes as shown in 
Fig.~\ref{Fig:Edge_states}(c).


\section{Experimental outlook}\label{Sec:exp_sys}
\begin{figure*}[!htb]
  \includegraphics[width=1.0\textwidth]{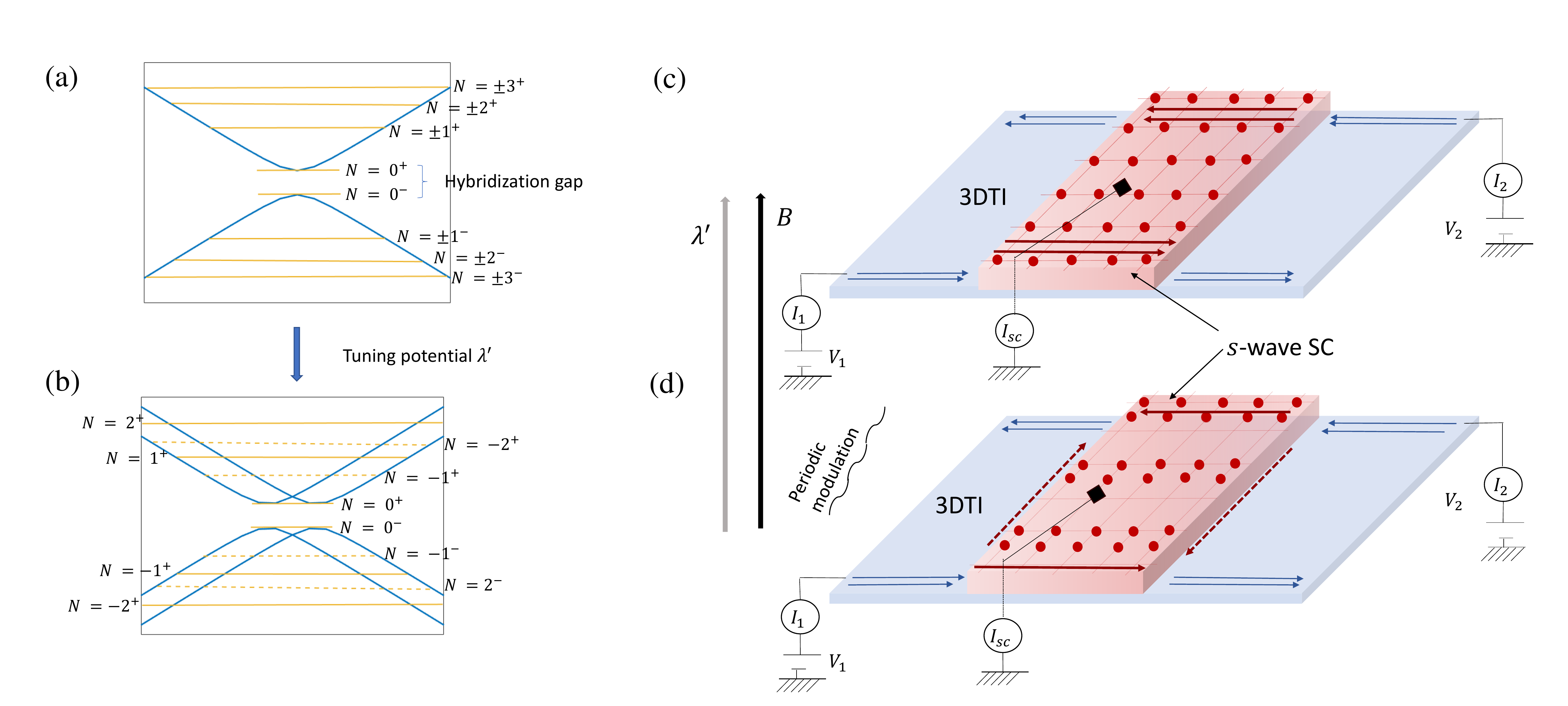}
  \caption{\label{Fig:Exp_setup}
  Experimental setup for topological superconductivity in the quantum Hall regime
  (a) Schematic of Landau levels of surface state of a thin film 3D Topological insulator (3DTI) with a hybridization gap between two layers. 
  The spectrum is doubly degenerate,
  (b) The Landau level degeneracy can be broken by a displacement field between the two layers leading to  
  normal state Landau level transitions with odd integer Hall conductance differences,
  (c) A quantum Hall system proximity coupled to a superconductor 
  in a $q=1$ vortex lattice phase,
  (d) periodic in plane field, the vortex lattice structure can be modified to $q = 2$ vortex lattice. An odd $\mathbb{Z}$ phase emerges in the central superconductor covered region.
   }
\end{figure*}

The relationship we have established between the vortex lattice configuration and the topological 
classification of superconducting states evolves from the factor of two difference the magnitudes 
of the electron and superconducting flux quanta,  and from magnetic translation group properties.
It is not dependent on the underlying zero-magnetic field electronic structure.  
We have focused on two-dimensional Dirac bands here because they lead to large Landau level separations at relatively weak magnetic fields, and therefore seem to have the best chance of being compatible with proximity
superconductivity in resolved Landau levels. 

Indeed, the first major experimental challenge in exploring this relationship lies in achieving Landau quantization in the regime where superconductivity still survives.
When the Landau level separation in a two-dimensional system is comparable to other energy scales, like the proximitized superconducting gap, we can expect edge states to have relatively short localization lengths and to be experimentally accessible via
transport experiments.  
This circumstance is generally referred to as the quantum Hall regime.  
Dirac systems with large Fermi velocities are ideal for achieving the quantum Hall regime without 
destroying superconductivity.  
For monolayer graphene ($v_F\sim 10^6 m/s$) and the  Dirac surface states of a three dimensional (3D) topological insulators ($v_F\sim 5\times 10^5 m/s$), the largest Landau level gap already exceeds $10 meV$ at $B\sim 1T$. Moreover, superconductivity has recently been successfully
induced in both graphene~\cite{Heersche2007,Du2008,Allen2015,Bretheau2017} and topological insulators~\cite{Wang2012,Xu2014,Xu2015}.
 
The next challenge is to eliminate double degeneracies that the system might possess in the normal state, for example degeneracies associated with spin, since these degeneracies tend to favor even-$\mathbb{Z}$ topological phases. 
Monolayer graphene has spin and valley degeneracy and very weak intrinsic spin-orbit coupling. 
Because of these, quantum Hall transitions in monolayer graphene occur in multiples
of four (\textit{i.e.} $\sigma_{xy} = e^2(4N+2)/h$) unless spin and valley symmetries are spontaneously broken~\cite{Nomura2006,Jiang2007,Checkelsky2008,Young2012}.
This makes monolayer graphene unfavorable for an odd-$\mathbb{Z}$ topological superconductivity phase in the quantum Hall regime.
The surface states of a 3D topological insulator thin film are 
effectively spinless due to strong spin-orbit coupling.  However there are two surface states, 
from top and bottom surfaces, and these supply an extra degeneracy.
A gate displacement field will induce a potential difference between top and bottom surfaces,
$\lambda'$, which lifts this final degeneracy and leads to single particle Landau level energies
\begin{align}
    & \xi^{\pm}_N = \pm \sqrt{\Big(\frac{\lambda'}{2}+S_N \epsilon_0\sqrt{|N|B}\, \Big)^2+\lambda^2}\, .
\end{align}
Here, $\lambda$ is the hybridization energy between top and bottom layer surface states
and $S_N=\pm$ is the surface-dependent Dirac chirality.   
The potential difference 
$\lambda'$ breaks the degeneracy of the $N^{th}$ Landau level by splitting it into $N^{+}$ and $N^{-}$ levels. 
For a typical 5-6 quintuple layers] thick 3D topological insulator, 
the layer hybridization energy $\lambda\sim 5-8$ meV allowing 
a value of $\lambda'$ on the same scale to induce clear Landau level 
separations at around $1T$ field, where thin film Nb-superconductor is well below its $H_{c2}$.

We propose an external magnetic field generalization,
illustrated schematically in  Fig.~\ref{Fig:Exp_setup} (c)-(d),  of the 
magnetized topological insulator experiment 
of He \textit{et. al.}~\cite{He2017} in which a thin Nb film
was deposited on a a thin film of a Cr doped 3D topological insulator, in our proposal the topological insulator is not doped.
In the case of interest here, we propose  the entire system is placed
under a perpendicular magnetic field, such that the surface states of the 3D topological insulator are in the QH regime. 
The edge states indicated in Fig.~\ref{Fig:Exp_setup} (c), correspond to BdG doubling of the single edge channel expected when one surface state 
conduction is below the Fermi level.
The two edge state channel configuration in the bare topological insulator region is equivalent to one QH edge state in the absence 
of pairing.  In the region covered with the superconductor, there are two corresponding Bogoliubov edge states.
In Fig.~\ref{Fig:Exp_setup} (d) we imagine that the vortex lattice configuration has been manipulated 
by spatially varying magnetic field strength, the thickness of the superconducting film, or any other property 
in a manner that is commensurate with the natural vortex lattice so as to convert from a $q=1$ 
vortex lattice to a $q=2$ vortex lattice. 
For example a weak in-plane magnetic field, which is made to go through a nearby (but not proximity coupled to the system) bulk superconductor to achieve a spatially periodic magnetic field profile can be used to control the vortex lattice structure in the active system.
This is equivalent to tuning $\Delta_0-\Delta_1$, as studied theoretically in our model calculations.  
Beyond a critical value of $\Delta_0-\Delta_1$, the surface that is covered by superconductor can 
host only one Bogoliubov edge state, as shown in Fig.~\ref{Fig:Exp_setup} (d).
In this configuration, one of the two edge states from the BdG doubled space in the bare topological insulator region is reflected from the 
superconducting region and the other is transmitted.
As argued by Chung \textit{et. al.}~\cite{Chung2011}, 
such a reduction should lead to a half-quantized 
longitudinal conductance plateau in two 
terminal transport measurement.  When the Majorana mode is induced by 
magnetization reversal, it has been argued~\cite{Ji2018,Huang2018} that a similar 
half-quantized conduction can result simply from strong disorder.  
In the present case, vortex lattice manipulation does not 
introduce additional disorder, in particular in the un-proximitized 
regions of the topological insulator.  
A clear half quantized conductance plateau measured in this way would 
therefore be a compelling signature of a Majorana edge mode and more importantly an effective 2D spinless $p\pm ip$ superconducting phase.

\section{Discussion and Conclusion}\label{Sec:discussion}

It is well known that vortices in effective $p$-wave superconductors host bound Majorana 
zero-modes~\cite{Cheng2009,Biswas2013,Liu2015,Murray2015}.  In an external magnetic field a vortex lattice~\cite{Liu2015,Murray2015} forms and the Majorana modes start to overlap to form low energy Majorana mini-bands, 
that are initially well separated from higher energy excitations of the superconductor.  
The limit considered here is reached at still
stronger magnetic fields, at which the vortices overlap substantially and the vortex core spectrum is not well separated from other excitations.
In this limit, most of our results generally apply for Landau levels emerging from an ordinary 2DEG as well as Dirac electrons.
This is important, because in the weak field limit, the ordinary 2DEG coupled to an $s$-wave superconductor is not an effective $p$-wave superconductor. 
We have established a relationship between a classification we introduce for vortex lattice structures,
and the topological classifications of superconducting phases which applies in this quantum Hall regime.
Even though Majorana mini-bands can no longer be distinguished in the bulk, Majorana edge channel modes are present when the superconducting state has an odd $\mathbb{Z}$ topological index.  

Achieving superconductivity in the Landau level regime is becoming more commonplace~\cite{Rickhaus2012,Shalom2015,Amet2016}, 
but it is still a challenge. 
There are theoretical predictions of superconductivity in two dimensions beyond the semiclassical $H_{c2}$~\cite{Tesuanovic1989,Tesanovic1991,Akera1991,MacDonald1992}, with $T_c$ increasing with magnetic field. Theoretically, such a re-entrant superconducting phase is possible because of very high density of state in the isolated Landau level regime of the parent superconductor with effective attractive interactions. However, strong field re-entrant phase is not observed experimentally, possibly due to Pauli breakdown and disorder. 
In our experimental proposal, the Landau level gap of the normal part is large enough even at moderate magnetic field that the parent superconductor can exist in vortex lattice phase well below its $H_{c2}$.  
Although most easily probed by edge-sensitive transport experiments, the relationship between vortex lattices and topological classification has a bulk origin, related ultimately to the difference between the electron and Cooper-pair magnetic flux quanta. One important future direction of work is the effect of disordered vortex lattices, and possibility of using disorder to induce odd-$\mathbb{Z}$ topological index in the TSC regime.

\acknowledgments
This work was supported by Welch foundation under the Grant Number TBF1473 and  Army Research Office through the MURI program by the Grant Number W911NF-16-1-0472.

\textit{Note added}- After completion of this work, a recent related work appeared~\cite{Mishmash2019} focusing on Majorana zero modes trapped in the vortex core in single Landau level limit.

\appendix

\section{Pairing matrix elements in Landau level basis}\label{app:Pairing_LL}
In this section we derive expression for pairing matrix element in Landau level basis. The evaluation involves calculating the integral in Eq.~\ref{Eq:Pairing_matrix_element_GC}. For the sake of generality, the pair potential $\Delta(\bm{r})$ when solved self consistently in the parent superconductor, takes the form of sum of different COM channels $j$~\cite{Norman1995}:
\begin{align}
     & \Delta(\bm{r}) = \sum_{j,t} \Delta_{j,t} \phi_{j,\sqrt{2}\,  t a_y} (\sqrt{2}\bm{r})\, .
\label{Eq:app_Pair_potential_general_channel}      
\end{align}
Notice that the pair potential used in main text is simply just keeping $j = 0$ channel of the above form.
The pairing matrix element between 2DEG-Landau levels as shown in Eq.~\ref{Eq:Pairing_matrix_element_GC}
\begin{align}
     \mathcal{G}^{N,M}_{Y,Y'} = \sum_{t,j} \int d\bm{r} \Delta_t \phi_{j,\sqrt{2}t a_y}(\sqrt{2}\bm{r}) \phi^{\ast}_{N,Y}(\bm{r})\bar{\phi}_{M,Y'}(\bm{r})\, ,
\label{Eq:app_Pairing_matrix_2DEG_int}     
\end{align}
after substituting the pair potential, the matrix element involves integrals over product of three Landau level wavefunctions.
Since $\bar{\phi}_{M,Y'} = \phi^{\ast}_{M,-Y'}$, transforming $Y' \rightarrow -Y'$(since it is just a dummy variable in current form), and using the transformation to the COM and relative coordinate systems, the identity~\cite{MacDonald1992} 
\begin{align}
     &\phi_{N,Y}(\bm{r_1})\phi_{M,Y'}(\bm{r_2}) = \sum_j \mathcal{B}^{N,M}_j \phi^{R}_{j,Y_c}([\bm{r_1}+\bm{r_2}]/2)\notag\\
     &\hspace{3cm}\times\phi^r_{N+M-j,Y_r}(\bm{r_1}-\bm{r_2})\, ,
\label{Eq:app_COM_r_LL_trans} 
\end{align}
can be used. Here the transformation matrix $\mathcal{B}^{N,M}_j$ is defined in Eq.~\ref{Eq:app_transformation_identitites}, $Y_c$, and $Y_r$ are COM and relative guiding centers respectively. The  transformation to COM and relative coordinates simplifies pairing matrix elements following the steps:
\begin{align}
    &\mathcal{G}^{N,M}_{Y,Y'} = \sum_t\sum_{j,j'} [\Delta_{t,j} \mathcal{B}^{N,M}_{j'}  \phi^{r\ast}_{N+M-j',Y_r}(0)\notag\\
    &\hspace{2.5cm}\times\int d\bm{r}  \phi_{j,\sqrt{2}ta_y}(\sqrt{2}\bm{r}) \phi^{R\ast}_{j',Y_c}(\bm{r})] \notag\\
    &\hspace{0.8cm} = \sum_t\sum_{j,j'} [\Delta_{t,j}\mathcal{B}^{N,M}_{j'} \varphi_{N+M-j'} \biggl (-\frac{Y_r}{\sqrt{2}\ell}\biggr )\frac{\delta_{Y_c,ta_y}}{\sqrt{L_x}}\notag\\
    &\hspace{1.5cm}\times \int dy\varphi_{j}\biggl (\frac{\sqrt{2}}{\ell}(y-ta_y)\biggr )\varphi_{j'}\biggl (\frac{\sqrt{2}}{\ell}(y-Y_c)\biggr )]\, .
\label{Eq:app_Pairing_matrix_sim_steps}    
\end{align}
Here $Y_c = (Y+Y')/2$ and $Y_r =  Y-Y'$. The pairing matrix element is then
\begin{align}
     &\mathcal{G}^{N,M}_{Y,Y'} = \sum_{t,j} \Delta_{t,j} \mathcal{B}^{N,M}_j \varphi_{N+M}\biggl (-\frac{Y_r}{\sqrt{2}\ell}\biggr )\frac{\delta_{Y+Y',2ta_y}}{\sqrt{L_x}}\, ,
\label{Eq:app_Pairing_matrix_GC_2DEG}     
\end{align}
which after sum over $t$ and using Eq.~\ref{Eq:phi_chi} substituting $\chi_{|N|+|M|-j}$ for $\varphi_{|N|+|M|-j}$ is the same as written in Eq.~\ref{Eq:app_pairing_channel} of the main text.
When the guiding centers are represented by integer stripe index
and a continuum as described in main text, the COM and relative guiding centers are described as,
\begin{subequations}
\label{Eq:app_strip_cont}
\begin{align}
     & Y_c = (s+s') a_y/2+(k_x+k'_x)\ell^2/2\, , \\
     & Y_r = (s-s') a_y + (k_x-k'_x)\ell^2\, ,
\end{align}
\end{subequations}
and the condition $\delta_{Y_c,ta_y}$, gives $k_x = -k_x'$, $s+s' = 2t$.
The above matrix elements are calculated in the ordinary 2DEG Landau level basis for the compactness of the expression, however, all the numerical calculation in our main text are performed in Dirac Landau level basis. The transformation of the above algebra to Dirac Landau levels $N$ and $M$ follows simply as,
\begin{align}
    &\mathcal{F}^{N,M}_{Y,Y'} = \sum_{t}\int d\bm{r} \Delta_t \phi_{0,\sqrt{2}ta_y}(\sqrt{2}\bm{r})\psi^{\ast}_{N,Y}(\bm{r})\bar{\psi}_{M,Y'}(\bm{r})\notag\\
    &\hspace{0.9cm}=\mathcal{N}_N\mathcal{N}_M\sum_{t}\int d\bm{r} \Delta_t \phi_{0,\sqrt{2}ta_y}(\sqrt{2}\bm{r})\notag\\
    &\hspace{3cm}\times[S_N\phi^{\ast}_{|N|-1,Y}(\bm{r})\bar{\phi}_{|M|,Y'}(\bm{r})\notag\\
    &\hspace{3cm}-S_M\phi^{\ast}_{|N|,Y}(\bm{r})\bar{\phi}_{|M|-1,Y'}(\bm{r})]\, ,
\label{Eq:app_Pairing_matrix_GC_Dirac_step}    
\end{align}
which simplifies to
\begin{align}
    &\mathcal{F}^{N,M}_{Y,Y'} =  \sum_t \Delta_t \mathcal{D}^{N,M}_0 \varphi_{N+M}\biggl (-\frac{Y_r}{\sqrt{2}\ell}\biggr )\frac{\delta_{Y+Y',2ta_y}}{\sqrt{L_x}}\, .
\label{Eq:app_pairing_matrix_GC_Dirac}    
\end{align}
Above only $j = 0$ pairing channel is retained.
Here $\mathcal{D}^{N,M}_j$ is defined in Eq.~\ref{Eq:Unitary_Dirac_LL} of main text. Using the integer stripe index and continuum label representation for the guiding center in Eq.~\ref{Eq:app_pairing_matrix_GC_Dirac}, we can obtain Eq.~\ref{Eq:Pairing_matrix_element} of main text.

\section{Higher vorticity vortex lattices}\label{app:higher_vor}
In this section, we consider cases of vortex lattices with higher order vorticity of each vortex. For simplicity, we discuss the cases of vorticity two and three and a geometric vortex lattice associated with transnational symmetry defined by $q = 1$ (See the discussion in Sec.~\ref{Sec:VL_states}), as shown in the Fig.~\ref{Fig:Vortex_lattice} (a), (b). 
Owning to the higher vorticity of each vortex, the simplest geometric vortex lattice structures can have even or odd superconducting flux through the vortex lattice unit cell, depending on the vorticity.  
We demonstrate that our main result that vortex lattices with odd superconducting flux per unit cell can not host odd-$\mathbb{Z}$ TSC phase, still applies.

The Symmetric gauge is a more convenient choice to write down wavefunctions with periodic arrangement of zeros of desired multiplicity.
However, the Landau gauge is more convenient choice for us to analyze the vortex lattice properties. 
Hence, we first write down some transformation properties between the two gauges.
For magnetic field $\bm{B} = B\hat{z}$, the Landau gauge and symmetric gauge vector potentials are respectively,
\begin{subequations}\label{Eq:app_Vec_pot}
\begin{align}
     & A_L = B(-y,\, 0)\, ,\\
     & A_S = \frac{B}{2}(-y,\, x)\, ,
\end{align}
\end{subequations}
such that
\begin{align}\label{Eq:app_vec_pot_rel}
    & A_S = A_L+\nabla \chi\, ,
\end{align}
where,
\begin{align}\label{Eq:app_grad_pot}
    & \chi = \frac{Bxy}{2}\, .
\end{align}
The wavefunctions of any charged particle with charge $\tilde{e}$ in the two gauges are related by the gauge transformation:
\begin{align}\label{eq:app_SL_trans}
    & \psi_S = U(\tilde{e},B) \,\psi_L\, ,\quad U(\tilde{e},B) = \textup{e}^{i\tilde{e}\chi/\hbar}\, .
\end{align}

The pair potential with $q = 1$ geometric vortex lattice and unit vorticity can be written in Landau gauge as(see Eq.~\ref{eq:Delta})
\begin{align}\label{eq:app_LG_pair}
     & \Delta_{1L}(\bm{r}) = \Delta_0\sum^{\infty}_{t= -\infty} \textup{e}^{i\lambda t^2}\textup{e}^{2ita_y x/\ell^2}\,\textup{e}^{-\frac{1}{\ell^2}(y-ta_y)^2}\, .
\end{align}
This wave function can expressed in terms of the Jacobi theta function~\cite{Mumford1983}:
\begin{align}\label{eq:app_LG_theta}
     & \Delta_{1L}(\bm{r}) = \Delta_0\textup{e}^{-y^2/\ell^2} \theta_3\biggl (\frac{a_y}{\ell^2}(x-iy),\frac{\lambda}{\pi}+i\frac{a_y}{a_x}\biggr )\, ,
\end{align}
where the Jacobi theta function is a quasi-periodic analytic function with simple zeros, defined over two complex valued arguments $z, \tau$:
\begin{align}\label{eq:app_Jacobi3}
     & \theta_3(z,\tau) = \sum_t \textup{e}^{i\pi \tau t^2}\,\textup{e}^{2i z t}\, .
\end{align}

We use numbered subscripts to denote the vorticity and the subscript `$S$' and `$L$' to denote symmetric and Landau gauge respectively.
The symmetric gauge representation of the above pair potential can be obtained using the transformation in Eq.~\ref{eq:app_SL_trans},
\begin{align}\label{eq:app_symm_pair}
    & \Delta_{1S}(\bm{r}) = \textup{e}^{-2ieBxy/(2\hbar)}\,\Delta_{1L}(\bm{r})\, .
\end{align}
Notice the use of $\tilde{e} = -2e$ for the Cooper pair of electrons.
The above expression can be caste into more familiar form of the symmetric gauge wavefunctions by substituting the complex variables $z = x+iy$ and $\bar{z} = x-iy$,
\begin{align}\label{eq:app_symm_pair_z}
    & \Delta_{1S}(\bm{r}) = \Delta_0\,\textup{e}^{-\frac{|z|^2}{2\ell^2}}\,\textup{e}^{\frac{\bar{z}^2}{2\ell^2}}\,\sum^{\infty}_{t = -\infty} \textup{e}^{\biggl (t^2(i\lambda-a_y/a_x)+\frac{2ita_y\bar{z}}{\ell^2}\biggr )}\, .
\end{align}
Notice, since above wavefunction is the pair wavefunction and the associated COM magnetic length $\ell_C = \ell/\sqrt{2}$, the form of the above wavefunction
\begin{align}\label{eq:app_symm_LLL}
    & \Delta_{1S}(\bm{r}) = \Delta_0\,f(\bar{z})\,\textup{e}^{-\frac{|z|^2}{4\ell^2_C}}\, ,
\end{align}
where 
\begin{align}\label{eq:app_analytic}
     & f(\bar{z}) = \textup{e}^{\frac{\bar{z}^2}{4\ell^2_C}}\,\sum^{\infty}_{t = -\infty} \textup{e}^{\biggl (t^2(i\lambda-a_y/a_x)+\frac{ita_y\bar{z}}{\ell^2_C}\biggr )}\, ,
\end{align}
is an analytic function of $\bar{z}$, clearly shows that $\Delta_{1S}(\bm{r})$ lies in the lowest symmetric gauge Landau level. Further, since the sum over $t$ part is a Jacobi theta function, it can be rewritten using the product representation of the theta function~\cite{Mumford1983}
\begin{align}\label{eq:app_theta_mul}
    & \theta_3\biggl (\frac{a_y\bar{z}}{\ell^2},\, \frac{\lambda}{\pi}+i\frac{a_y}{a_x}\biggr ) = \Pi^{\infty}_{m = 1}(1-\textup{e}^{2im\pi\biggl (\frac{\lambda}{\pi}+i\frac{a_y}{a_x}\biggr )})\notag\\
    &\hspace{3cm}\times(1+\textup{e}^{i(2m-1)\pi\biggl (\frac{\lambda}{\pi}+i\frac{a_y}{a_x}\biggr )+\frac{2ia_y\bar{z}}{\ell^2}})\notag\\
    &\hspace{3cm}\times(1+\textup{e}^{i(2m-1)\pi\biggl (\frac{\lambda}{\pi}+i\frac{a_y}{a_x}\biggr )-\frac{2ia_y\bar{z}}{\ell^2}})\, .
\end{align}
One can see that the zeros of the pair wavefunctions (vortices) occur at
\begin{align}\label{eq:app_zeros_theta}
     & \bar{z} = \pm [(2n+1)-(2m-1)\lambda/\pi]\frac{a_x}{2}\mp i (2m-1)\frac{a_y}{2}\, ,
\end{align}
where $n\in \mathcal{Z}$ and $m\in \mathcal{Z}^+$.

To construct the pair potential with vortices of vorticity $n$ in the symmetric gauge, as a first step we simply raise the pair potential in Eq.~\ref{eq:app_symm_pair} to power $n$ upto a normalization factor $\mathcal{N}$. The $n^{th}$ power of the analytic part $f(\bar{z})$ ensures all the zeros have $n^{th}$ order multiplicity (have vorticity $n$),
\begin{align}\label{eq:app_nth_vor}
      & \Delta_{n,S}(\bm{r}) = \mathcal{N}\Delta^n_{1S}(\bm{r})\notag\\
      & \hspace{1.2cm} = \mathcal{N}\,\textup{e}^{-\frac{n|z|^2}{2\ell^2}}\,f^n(\bar{z})\, .
\end{align}
However, this increases the density of the vortices by a factor of $n$, which is physically only possible if the magnetic flux is increased by a factor of $n$. Since
\begin{align}\label{eq:app_mag_n_rel}
    & \textup{e}^{-\frac{n|z|^2}{2\ell^2}} = \textup{e}^{-\frac{|z|^2}{2\tilde{\ell}^2}}\, , \quad \tilde{\ell} = \frac{\ell}{\sqrt{n}}\, .
\end{align}
We can interpret the new pair potential $\Delta_{n,s}$ as the pair potential in the magnetic field increased by a factor of $n$. This way, the geometric vortex lattice remains unchanged, however the  vorticity of each vortex is multiplied by the factor of $n$ as the magnetic field is increased to $n$-times. 
All the other changes associated with increased magnetic field can be included in the change in the normalization factor $\mathcal{N}$.
While calculating the matrix elements of the pair potential, one needs to be careful in using single particle Landau level states in magnetic field increased by factor of $n$. 
Consequently, the Landau gauge representation of the higher vorticity vortex lattice  follows:
\begin{align}\label{eq:app_L_gauge_trans}
    & \Delta_{n,L}(\bm{r}) = U^{\dagger}(nB)\Delta^n(n,S)\notag\\
    &\hspace{1.25cm} = U^{\dagger}(nB) U^n(B)\Delta^n_{1,L}\notag\\
    &\hspace{1.25cm} = \Delta^n_{1,L}\, .
\end{align}

It is important to note that the above only works because the pair potential lies in the lowest Landau level. Higher Landau level in symmetric gauge have differential operators, hence the gauge transformation and wavefucntion do not commute. 
The same algebra is manifested in Landau gauge via Hermite polynomials. 
By increasing the power of the Landau gauge wavefunction in higher Landau level, the Hermite polynomials give higher order Hermite polynomial. 
Hence the wavefunction does not remain in the Hilbert space of the same Landau level.

To calculate the BdG matrix elements of the pair wavefunction, we need to simplify the higher order  theta function to the first order theta functions.
Below we will use two additional Jacobi theta functions defined by
\begin{subequations}
\begin{align}
    & \theta_1(z,\tau) = -i\textup{e}^{\frac{i\pi\tau}{4}}\sum^{\infty}_{t = -\infty} (-)^t\textup{e}^{i\pi\tau(t^2+t)}\textup{e}^{(2t+1)iz}\, , \\
    & \theta_2(z,\tau) = \textup{e}^{\frac{i\pi\tau}{4}}\sum^{\infty}_{t = -\infty} \textup{e}^{i\pi\tau(t^2+t)}\textup{e}^{(2t+1)iz}\, .
\end{align}
\end{subequations}
Now, we list some important identities involving Jacobi theta functions which are useful for the purpose of calculating BdG matrix elements~\cite{Mumford1983,He2018}:
\begin{subequations}
\begin{align}
    & \theta^2_1(z,\tau) = \theta_2(0,2\tau)\theta_3(z,\tau/2)-\theta_2(2z,2\tau)\theta_3(0,\tau/2)\, ,\\
    & \theta^3_1(z,\tau) = \theta_1(z,\tau/3)\frac{\theta^3_1(0,\tau)}{\theta_1(0,\tau/3)-\frac{\theta_1(0,\tau)\theta'_1(0,\tau/3)}{\theta'_1(0,\tau)}}\, ,\\
    & \theta_1(z+\frac{\pi}{2}(\tau+1),\tau) = \theta_2(z+\frac{\pi\tau}{2},\tau)\, ,\\
    & \theta_3(z,\tau) = \textup{e}^{\frac{i\pi\tau}{4}}\,\textup{e}^{iz}\,\theta_2(z+\frac{\pi\tau}{2},\tau)\, \notag\\
    &\hspace{1.1cm} = \textup{e}^{\frac{i\pi\tau}{4}}\,\textup{e}^{iz}\, \theta_1(z+\frac{\pi}{2}(\tau+1),\tau)\, .
\end{align}
\end{subequations}
Using the above identities, the square of the Jacobi theta function is 
\begin{align}
    & \theta^2_3(z,\tau) = \textup{e}^{i\pi\tau/2}\,\textup{e}^{2iz} \,[\alpha_{\tau}\theta_3(z+\frac{\pi}{2}(\tau+1),\tau/2)\notag\\
    &\hspace{2cm}-\beta_{\tau}\textup{e}^{-2iz}\textup{e}^{-i\pi\tau/2}\theta_3(2z+\pi,2\tau)]\, ,
\end{align}
where
\begin{align}
     & \alpha_{\tau} = \theta_2(0,2\tau)\, , \quad \beta_{\tau} = \theta_3(0,\tau/2)\, . 
\end{align}

Similarly,
\begin{align}
    & \theta^3_3(z,\tau) = \textup{e}^{2iz}\,\textup{e}^{i\pi\tau/3}\,\gamma_{\tau}\,\theta_3(z+\pi\tau/3,\tau/3)\, ,
\end{align}
where
\begin{align}
    & \gamma_{\tau} = \frac{\theta^3_1(0,\tau)}{\theta_1(0,\tau/3)-\frac{\theta_1(0,\tau)\theta'_1(0,\tau/3)}{\theta'_1(0,\tau)}}\, .
\end{align}

Next, we calculate the matrix elements of the pair potential in the Landau quantized single electron basis for the case of vorticity two and three.

\underline{Vorticity two}- If each vortex has vorticity two, the pair potential in the Landau gauge takes the form
\begin{widetext}
\begin{align}
     & \Delta_{2L}(\bm{r}) = \mathcal{N}\Delta^2_0 \textup{e}^{-y^2/\tilde{\ell}^2}\,\theta^2_3(w,\tau) \notag\\
     &\hspace{1.25cm} = \mathcal{N}\Delta^2_0 \textup{e}^{-y^2/\tilde{\ell}^2}\,\textup{e}^{2iw}\textup{e}^{i\pi\tau/2}[\alpha_{\tau}\theta_3(w+\frac{\pi}{2}(\tau+1),\tau/2)-\beta_{\tau}\textup{e}^{-2iw}\textup{e}^{-i\pi\tau/2}\theta_3(2w+\pi,2\tau)]\, \notag\\
     &\hspace{1.25cm} = -\mathcal{N}\Delta^2_0[\alpha_{\tau}\sum_t (-)^t \textup{e}^{i\lambda t^2/2}\textup{e}^{ita_yx/\ell^2}\textup{e}^{-\frac{1}{\ell^2}(y-ta_y/2)^2} +\beta_{\tau}\sum_t \textup{e}^{2 i\lambda t^2}\textup{e}^{2ita_yx/\ell^2}\textup{e}^{-\frac{1}{\ell^2}(y-ta_y)^2} ]
\end{align}
\end{widetext}
where,
\begin{align}
    & w = \frac{a_y}{2\tilde{\ell}^2}(x-iy)\, ,\quad \tau = \frac{\lambda}{\pi}+i\frac{a_y}{a_x}\, ,
\end{align}
and,
\begin{align}
   & a_x a_y = 2\pi\tilde{\ell}^2\, , \quad \tilde{\ell} = \ell/\sqrt{2}\, .
\end{align}
calculating the matrix element of the pair potential $\Delta_{2L}(\bm{r})$ in the ordinary 2DEG Landau level basis, where the guiding centers are $Y = sa_y+k_x\ell^2$ and $Y' = s'a_y-k_x\ell^2$, we obtain
\begin{widetext}
\begin{align}
   & \mathcal{G}^{N,M}_{s,s'} = -\frac{\mathcal{N}}{\sqrt{L_x}}\Delta^2_0\varphi_{N+M}\biggl (-\frac{Y_r}{\sqrt{2}\tilde{\ell}}\biggr ) \mathcal{B}^{N,M}_0[\sum_t  ((\alpha_{\tau}+\beta_{\tau}) \textup{e}^{2i\lambda t^2}\delta_{s+s',2t}-\alpha_{\tau}\textup{e}^{i\lambda (2t+1)^2/2}\delta_{s+s',2t+1})]\, .
\end{align}
\end{widetext}
The first term in the above matrix element couples odd integers with odd integers and even integers with even integers ($\delta_{s+s'} = 2t$ term), while the second term above couples odd integer index with even integer index.
If one tries to diagonalize the system in the representation of odd stripe index and even stripe index, as done in the main text, the non-zero off diagonal coupling leads to avoided crossings and breaks the double degeneracy.
Hence, it can allow situations with odd-$\mathbb{Z}$ TSC phase.
This agrees with our main result, since for $q = 1$ geometric arrangements, vortices of vorticity two have two superconducting fluxes per vortex lattice unit cell.

\underline{Vorticity three}- 
If each vortex has vorticity three, the pair potential in Landau gauge takes the form
\begin{align}
     & \Delta_{3L}(\bm{r}) = \mathcal{N}\Delta^3_0 \textup{e}^{-y^2/\tilde{\ell}^2}\,\theta^3_3(w,\tau) \notag\\
     &\hspace{1.25cm} = \mathcal{N}\Delta^3_0 \textup{e}^{-y^2/\tilde{\ell}^2}\,\textup{e}^{2iw}\textup{e}^{i\pi\tau/3}\gamma_{\tau}\, \theta_3(w+\pi\tau/3,\tau/3)\, \notag\\
     &\hspace{1.25cm} = \mathcal{N}\Delta^3_0\gamma_{\tau} \sum_t \textup{e}^{i\lambda t^2/3} \textup{e}^{2ita_yx/(3\ell^2)}\textup{e}^{-\frac{1}{\ell^2}(y-ta_y/3)^2}\, ,
\end{align}
where,
\begin{align}
    & w = \frac{a_y}{3\tilde{\ell}^2}(x-iy)\, ,\quad \tau = \frac{\lambda}{\pi}+i\frac{a_y}{a_x}\, ,
\end{align}
and,
\begin{align}
   & a_x a_y = 3\pi\tilde{\ell}^2\, , \quad \tilde{\ell} = \ell/\sqrt{3}\, .
\end{align}
calculating the matrix element of the pair potential $\Delta_{3L}(\bm{r})$ in the ordinary 2DEG Landau level basis, we obtain
\begin{align}
   & \mathcal{G}^{N,M}_{s,s'} = \frac{\mathcal{N}}{\sqrt{L_x}}\Delta^3_0\varphi_{N+M}\biggl (-\frac{Y_r}{\sqrt{2}\ell}\biggr ) \mathcal{B}^{N,M}_0\notag\\
   &\hspace{1.5cm}\times\sum_t [\gamma_{\tau} \textup{e}^{i\lambda t^2/3}\delta_{2(s+s'),3t}]\, .
\end{align}
The above expression only allows pairing between guiding centers with both even or both odd integer index labels.
Hence, the system again decouples into two degenerate subsystems and can only allow even-$\mathbb{Z}$ TSC phase.
Since, vorticity three vortex lattice has odd flux per vortex lattice unit cell, that is exactly what is expected from our main results.

\section{Vortex lattice symmetry and magnetic Bloch states}\label{app:Magnetic_BLoch}
When the superconductor has vortex lattice symmetry, such that $|\Delta_{t+q}| = |\Delta_t|$, we first write the integer index $s$ of the guiding center as $pq+m$, such that
\begin{subequations}
\label{Eq:app_GC_VL_symmetry_general}
\begin{align}
      & Y = (pq + m) a_y + k_x\ell^2\, , \\
      & Y' = (p'q + m') a_y - k_x\ell^2\, .
\end{align}
\end{subequations}
Then the pairing matrix element coupling guiding centers $Y$ and $Y'$ in this representation is given by
\begin{align}
     & \mathcal{F}^{N,M}_{\{p,m\};\{p',m'\}}(k_x) = \sum_{t,n}\Delta_{tq+n}\delta_{p+p',2t}\delta_{m+m',2n}\notag\\
     &\hspace{0.5cm}\times\chi_{|N|+|M|-1}([q(p-p')+(m-m')]a_y+2k_x\ell^2)\, .
\label{Eq:app_Pairing_GC_function}     
\end{align}
In this representation the BdG matrix equation takes form (only the upper block shown, lower block follows trivially):
\begin{widetext}
\begin{align}
     & \xi_N u^{\nu}_{N,p,m}(k_x)+\sum_{M, p',m'}\mathcal{F}^{N,M}_{\{p,m\};\{p',m'\}}(k_x) v^{\nu}_{M,p',m'}(k_x) = Eu^{\nu}_{N,p,m}(k_x) \notag\\
     &\implies \xi_N u^{\nu}_{N,p,m}(k_x)+\sum_{M} 
     \mathcal{F}^{N,M}_{\{p,m\};\{2t-p,2n-m\}}(k_x)v^{\nu}_{M,2t-p,2n-m}(k_x) = Eu^{\nu}_{N,p,m}(k_x)
\label{Eq:app_BdG_GC_non_zero_condition}     
\end{align}
\end{widetext}
To diagonalize the above matrix equations, we now transform to the magnetic Bloch states defined on the electron and hole part of the Landau level basis in the BdG equations respectively
\begin{widetext} 
\begin{subequations}
\label{Eq:app_magnetic_Bloch}
\begin{align}
    &\phi_{N,m,\bm{k}}(\bm{r}) = \sqrt{\frac{qa_y}{L_y}}\sum_t \textup{e}^{ik_y(qt+m)a_y}\textup{e}^{i\pi\lambda q t(t-1)/2} \textup{e}^{i(\pi\lambda m-\theta/2)t} \phi_{N,k_x\ell^2+(qt+m)a_y}(\bm{r})\, ,\\
    & \phi^{\ast}_{N,m,-\bm{k}}(\bm{r}) = \sqrt{\frac{qa_y}{L_y}}\sum_t \textup{e}^{ik_y(qt+m)a_y}\textup{e}^{-i\pi\lambda q t(t-1)/2} \textup{e}^{-i(\pi\lambda m-\theta/2)t} \phi^{\ast}_{N,-k_x\ell^2+(qt+m)a_y}(\bm{r})\, .
\end{align}
\end{subequations}
\end{widetext}
Now after expanding the eigenvector $\Psi^{\nu}_{m,\bm{k}}(\bm{r}) = (\sum_{N} u^{\nu}_{N,m,\bm{k}}\psi_{N,m,\bm{k}}(\bm{r}),\,\,\, \sum_{N} v^{\nu}_{N,m,\bm{k}}\psi^{\ast}_{N,m,-\bm{k}}(\bm{r}))^.$, the upper BdG block diagonalization equation
\begin{align}
\label{Eq:app_k_space_diag}
      & \xi_N u^{\nu}_{N,n}(\bm{k})+\sum_{M,m}\mathcal{F}^{N,M}_{n,m}(\bm{k}) v^{\nu}_{M,m}(k_x) = E(\bm{k})u^{\nu}_{N,n}(\bm{k}) \, ,
\end{align}      
is obtained for $n\in [0,..,q-1]$. Here the Fourier representation of the pairing matrix element is obtained as:
\begin{widetext}
\begin{align}
     &\mathcal{F}^{N,M}_{n,m}(\bm{k}) = \mathcal{D}^{N,M}_0\sum_{t}\Delta_{qt+(n+m)/2}\textup{e}^{-i\pi \lambda q t (t-1)}\textup{e}^{-i(2\pi\lambda n -\theta)t}\textup{e}^{ik_ya_y(2qt+n-m)}\chi_{|N|+|M|-1}[(2qt+n-m)a_y+2k_x\ell^2]\, .
\label{Eq:app_pairing_matrix_Dirac_k_space}     
\end{align}
\end{widetext}

For the purpose of the discussion in next section, we also explicitly write down the pairing matrix element for electrons in ordinary 2DEG Landau levels:
\begin{widetext}
\begin{align}
     &\mathcal{G}^{N,M}_{n,m}(\bm{k}) = \mathcal{B}^{N,M}_0\sum_{t}\Delta_{qt+(n+m)/2}\textup{e}^{-i\pi \lambda q t (t-1)}\textup{e}^{-i(2\pi\lambda n -\theta)t}\textup{e}^{ik_ya_y(2qt+n-m)}\chi_{N+M}[(2qt+n-m)a_y+2k_x\ell^2]\, .
\label{Eq:app_pairing_matrix_2DEG_k_space}     
\end{align}
\end{widetext}

\section{Superconductivity in Ordinary 2DEG Landau level}\label{app:2DEG}
In this section we extend the discussion in the main text to the ordinary 2DEG Landau levels and show the main results of inducing superconductivity in QH regime and the connection between vortex lattice structure and odd-$\mathbb{Z}$ classification of TSC is insensitive to this detail. Although the low field limit of the two cases is dramatically different. To convey the main point through simplified analysis, we consider only the $q = 1$ case.

\begin{figure}[!htb]
  \includegraphics[width=0.5\textwidth]{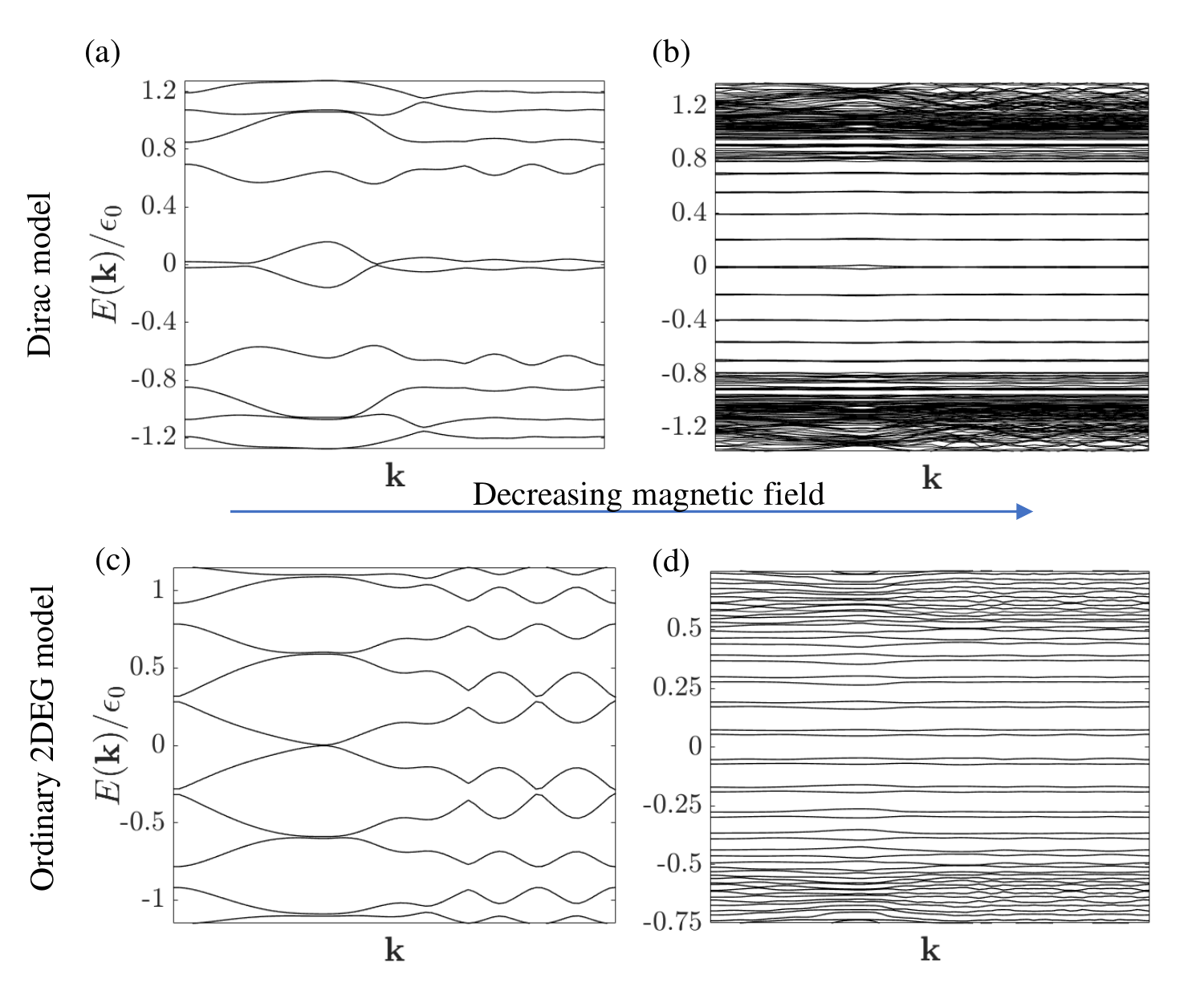}
  \caption{\label{Fig:Dirac_2DEG_comparison}
  Bulk BdG quasiparticle spectrum at strong (a) and (c) vs weak (b) and (d) field.
  For the Dirac model of the normal state, as the magnetic field is decreased the lowest energy vortex core bound states are at zero energy as shown in (b). This indicates the system is effective $p$-wave superconductor as predicted by Fu-Kane model~\cite{Fu2008}.
  In contrast when Landau levels of ordinary 2DEG are proximity coupled to $s$-wave superconductor, in the weak field limit, the lowest vortex core bound state is at finite energy as shown in (d). This is expected for  an $s$-wave superconductor.
   }
\end{figure}

To have a good comparative understanding of the two cases, first consider the single Landau level limit. When $\mu$ is exactly at Landau level energy, the spectrum:
\begin{align}
     & E(\bm{k}) = 
     \begin{cases}
     \pm |\mathcal{F}^{N,N}(\bm{k})|
     & \text{Dirac Landau level}\, ,\\
     \pm |\mathcal{D}^{N,N}(\bm{k})|
     & \text{Ordianry 2DEG}\, . 
     \end{cases}
\label{Eq:app_single_LL_spectrum}     
\end{align}
only differs through the order of the Hermite polynomial appearing in the pairing matrix element.
Now notice that, the single Landau level pairing matrix element for Dirac Landau level has odd Hermite polynomial of order $2|N|-1$, while for ordinary 2DEG, has even Hermite polynomial of order $2N$. Because of this, $\mathcal{F}^{N,N}(\bm{k}) = -\mathcal{F}^{N,N}(-\bm{k})$ and $\mathcal{D}^{N,N}(\bm{k}) = \mathcal{D}^{N,N}(-\bm{k})$. Hence, irrespective of type of vortex lattice, the BdG Hamiltonian in single Landau level is odd parity for Dirac case and even parity for ordinary 2DEG case. This means only the Dirac case in its single Landau level limit can be an effective $p$-wave system. Now focusing on only $\bm{k} = 0$ point, the Dirac case is gapless at that point with Dirac touching point, while ordinary 2DEG case is generally not. 

\begin{figure}[!htb]
  \includegraphics[width=0.5\textwidth]{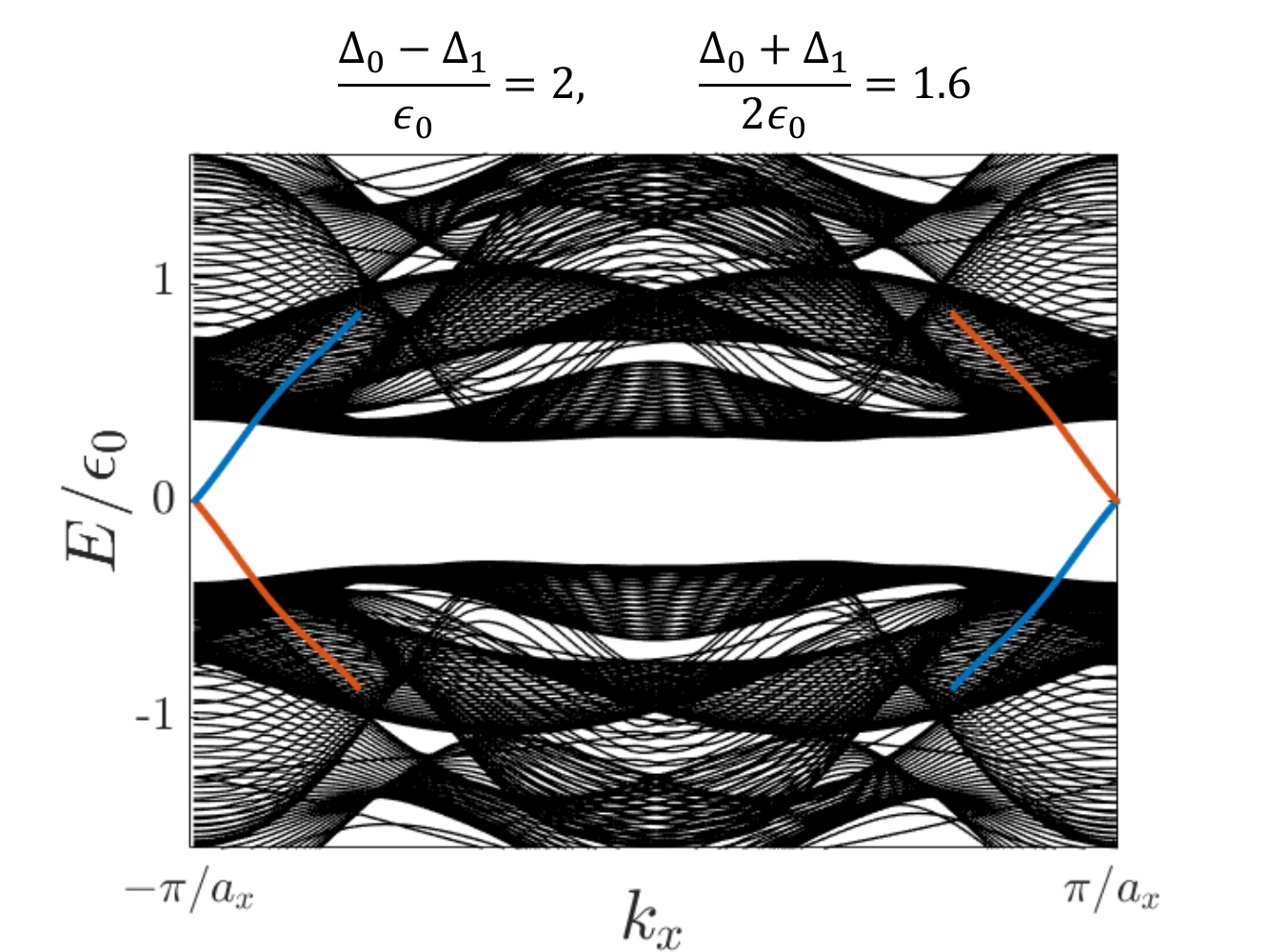}
  \caption{\label{Fig:2DEG_edge}
  A representative case of finite width spectrum (in the units of $\epsilon_0 = \hbar \omega_c$ of two ordinary 2DEG Landau levels proximity coupled to an $s$-wave superconductor. For clarity, we have removed the non topological edge state from the figure. The red and blue denote opposite edges. Hence, the system has exactly one chiral edge mode. The parameters are chosen fictitiously, such that Landau level gaps and superconducting pair potential amplitudes are comparable.
   }
\end{figure}

The discussion in single Landau limit, although brings out qualitative difference between effective pairing symmetry of Dirac Landau level case compared to 2DEG Landau level case, however under our construction single Landau level cannot drive topological phase changes in either scenario. If we turn to the opposite limit of weak magnetic field where many Landau levels contribute to pairing, the situation is again dramatically different for Dirac and ordinary 2DEG case. Dirac case with $\mathcal{T}$-symmetry proximitized to an $s$-wave superconductor is shown to be an effective $p$-wave superconductor~\cite{Fu2008}, while the ordinary 2DEG case is $s$-wave superconductor. This can be seen in the Fig.~\ref{Fig:Dirac_2DEG_comparison}(b) and (d). In the isolated vortex limit, which is relevant at weak magnetic field, an $s$-wave superconductor binds Caroli-de Gennes-Matricon states to the vortex core, which have energy levels $E_n\sim (n+1/2) \Delta^2/E_F$, where $\Delta$ is the superconducting gap and $E_F$ is the Fermi energy and $n$ is a positive integer~\cite{Caroli1964}. This can be seen by the finite energy of the lowest energy band in Fig.~\ref{Fig:Dirac_2DEG_comparison}(d). In contrast for the $p$-wave superconductor vortex core levels have energy $E_n\sim n\Delta^2/E_F$~\cite{Kopnin1991}. This is evident in the Fig.~\ref{Fig:Dirac_2DEG_comparison} (b) with the zero energy band of vortex core states. In terms of possibility of Majorana, this means the Dirac case can bind MZMs at the vortex core under suitable conditions, while the vortex core bound states for 2DEG case is always complex fermion beacuse of the associated  finite energy.

For the intermediate limit of a few Landau level, which is of focus in this work, the qualitative physics for Dirac and 2DEG is similar. In particular, in context of possibility of odd or even number of chiral Majorana edge modes, the vortex lattice structure is the main tuning knob. As shown in Fig~\ref{Fig:2DEG_edge} for a representative case of two landau level model of ordinary 2DEG, the odd number of chiral Majorana edge modes can be achieved by suitable tuning of Landau level gap, superconducting pair potential and vortex lattice structure.


\bibliography{bibliography}

\end{document}